\documentclass[aps,prb,twocolumn,showpacs,floatfix,twoside,superscriptaddress]{revtex4-2}
\usepackage{amssymb}
\usepackage{graphicx}
\usepackage{amsmath}
\usepackage{amsfonts}
\usepackage{bbold}
\usepackage{esint}
\usepackage{color}
\usepackage[dvipsnames]{xcolor}
\usepackage{float}
\usepackage{verbatim}
\usepackage[utf8]{inputenc}
\usepackage{verbatim}
\usepackage[colorlinks=true,citecolor=blue,linkcolor=blue,urlcolor=blue]{hyperref}
\usepackage[justification=centerlast,font={small}]{caption}
\usepackage{subcaption}
\usepackage{multirow}
\usepackage{placeins}

\newcommand{\lsum}{\sum\limits}
\newcommand{\lint}{\int\limits}

\DeclareMathOperator{\sign}{sign}

\newcommand{\BA}{\mathcal{A}}
\newcommand{\BB}{\mathcal{B}}

\newcommand{\BD}{\mathcal{D}}

\begin{document}

\title{Tomographic flow regime vs even-odd effect for the magnetotransport in the Corbino geometry}
\author{Grigory A. Starkov}
\affiliation{Institute for Theoretical Physics and Astrophysics,
University of W\"urzburg, D-97074 W\"urzburg, Germany}
\affiliation{W\"urzburg-Dresden Cluster of Excellence ctd.qmat, Germany}
\email{grigorii.starkov@uni-wuerzburg.de}

\date{\today}

\begin{abstract}
In two dimensions, the geometric constraints due to Pauli blocking and conservation laws lead to the even-odd effect exhibited by the electron-electron scattering lengths: electron-electron collisions are more efficient at relaxing the even angular harmonics of the distribution function than the odd ones. Inspired by a recent experiment on the magnetotransport in the Corbino disk geometry, we numerically analyze the electron flows in this geometry across all the regimes.

We predict a clear signature of the even-odd effect --- enhancement of the resistance sensitivity $\partial R/\partial(B^2)$ at small magnetic fields $B\rightarrow 0$. This enhancement is most prominent at the crossover from the ballistic to the tomographic regime, and gradually disappears when the temperature is further increased. Our estimates suggest that in the temperature range of the experiment, the effect should be small. This implies that the attribution of the anomalous scaling of the kinematic viscosity, that was observed in the experiment, to the even-odd effect might need more careful consideration.

As a side note, we show how the method of characteristics can be extended to treat the long-lived odd harmonics, which allows one to recast the linearized Boltzmann equation as a system of integral ones.

\end{abstract}

\maketitle

\section{Introduction}

Electron flows in metals are typically dominated by momentum-relaxing processes such as impurity scattering or electron-phonon scattering. The situation is different in two dimensions: the rate of electron-phonon collisions grows only linearly with temperature, while advances in device fabrication made it possible to produce ultra-clean samples.
As such, two-dimensional materials constitute a unique platform, where the hydrodynamic flow regime dominated by the momentum-conserving electron-electron collisions can be studied~\cite{Narozhny_2022}.

The hallmark feature of this flow regime is the Gurzhi effect~\cite{Gurzhi_1963,Gurzhi_1968} --- growth of the conductance with temperature. Normally, one expects that making electron-electron collisions stronger should lead to an increase of resistance. However, since they conserve momentum, they can not influence the resistance directly, and the non-trivial interplay with momentum-relaxing processes leads to this counter-intuitive effect.
The strength of the Gurzhi effect varies with geometry, and in some cases can even lead to superballistic flows~\cite{KrishnaKumar_2017, Sarypov_2025}, characterized by a larger conductance than in the ballistic limit. In principle, this can be used to decrease the dissipation of electric devices~\cite{Stern_2022, Huang_2024, Estrada-Alvarez_2025a, Estrada-Alvarez_2025}.

Due to energy-momentum conservation and Pauli blocking, the electron-electron collisions are restricted to a thin annulus in the vicinity of the Fermi surface.
In two dimensions, they are dominated by head-on collisions, in which the electrons meet with almost opposite momenta and scatter at large angles.
However, these processes can efficiently relax only the even part of the electron distribution function, $f_\mathrm{even}(\vec  k) = (f(\vec k) + f(-\vec k))/2$. For the odd part, $f_\mathrm{odd}(\vec  k) = (f(\vec k) - f(-\vec k))/2$, the relaxation is determined by the sub-leading channel, corresponding to processes in which the electrons approach each other with an arbitrary angle between their incoming momenta and undergo only small-angle deflections. This results in the even-odd effect --- vastly different relaxation rates for the even and odd parts of the distribution function~\cite{Gurzhi_1995, Ledwith_2019}.
To be precise, one can show (with logarithmic accuracy) in the limit of the degenerate Fermi gas $T/E_F\ll 1$ that the inverse scattering time for the even part scales as~\cite{Giuliani_1982,Zheng_1996, Nilsson_2025} 
\begin{equation}
\frac{1}{\tau^{(\mathrm{ee})}_e} \sim \frac{E_F}{h}\left(\frac{T}{E_F}\right)^2,\label{base-even-rate}
\end{equation}
while the inverse relaxation time for the odd part acquires an additional $(T/E_F)^2$ factor~\cite{Ledwith_2019a, Hofmann_2023, Nilsson_2025},
\begin{equation}
    \frac{1}{\tau^{(\mathrm{ee})}_o} \sim \frac{E_F}{h}\left(\frac{T}{E_F}\right)^4\label{base-odd-rate}.
\end{equation}
As a consequence, one expects the appearance of a new "tomographic" flow regime when the even part of the distribution function behaves hydrodynamically $L\gg v_F\tau^{(\mathrm{ee})}_e$ while the odd part still behaves ballistically $L\ll v_F\tau^{(\mathrm{ee})}_o$. Here, $L$ denotes a characteristic linear size of the system.

Elucidating the key signatures of the even-odd effect and the tomographic flow regime associated with it have been the focus of an ongoing theoretical~\cite{Kryhin_2023, Kryhin_2023a, Kryhin_2025, Hofmann_2023, Gran_2023, Hofmann_2024, Rostami_2025, Maki_2025, Maki_2026, Ben-Shachar_2025, Ben-Shachar_2025a, Ben-Shachar_2026, Raichev_2025, Musser_2026, Starkov_2026, Estrada-Alvarez_2025, Thuillier_2026, Farrell_2026} and experimental effort~\cite{Zeng_2024, Moiseenko_2025}.  In particular, the $k^{-5/3}$ scaling of the bulk transversal conductivity has been predicted~\cite{Kryhin_2025, Rostami_2025, Thuillier_2026} as the main characteristic of the tomographic regime, and the even-odd effect was demonstrated for the linewidths of the higher-order Cyclotron Resonance peaks.

The electron-electron collisions do not relax the odd part of the distribution function uniformly. To describe this effect, one should expand the nonequilibrium part of the distribution function in the angular eigenmodes of the linearized collision operator on the Fermi surface. If we assume the latter to be circular, this results in a Fourier series expansion,
\begin{equation}
    \delta f(\theta)\sim \eta_0^{(c)} + \sum_{m=1}^{+\infty} \left[\eta_m^{(c)}\cos{m\theta} + \eta_{m}^{(s)}\sin(m \theta)\right],
\end{equation}
where harmonics with even (odd) mode number $m$ correspond to the even (odd) part of the distribution function.
The inverse relaxation times $1/\tau_m$ of the eigenmodes are given by the eigenvalues of the linearized collision operator.
For the even harmonics, $1/\tau_m$ have only weak logarithmic dependence on $m$, which we neglect. For the odd modes,
the inverse relaxation time quickly grows with the mode number as $1/\tau_m = m^4/\tau^{(\mathrm{ee})}_o$ and saturates at the level of $1/\tau^{(\mathrm{ee})}_e$ for $m^\prime\sim \sqrt{E_F/T}$~\cite{Ledwith_2019a, Hofmann_2023, Nilsson_2025}. The number of the long-lived odd modes with anomalously slow relaxation is finite and quickly diminishes with the temperature. The even-odd effect disappears along with it.

Due to this reason, experimental observation of the pure tomographic regime with a huge number of the long-lived odd harmonics is actually rather hard: driving the even modes into the hydrodynamic regime requires high enough temperatures, at which not many of the long-lived harmonics may survive. On the other hand, the even-odd effect should be most prominent precisely when many slowly-relaxing modes are present, {\it i.e.\/}, in the ballistic regime $v_F\tau^{(\mathrm{ee})}_{e,o}\gg L$ and at the onset of the tomographic one $v_F\tau^{(\mathrm{ee})}_e\sim L$. These are the regimes, where 
one of the more striking consequences of the even-odd effect appears --- the growth of conductance at small temperatures.
Normally, the Gurzhi effect is observed when the electron-electron scattering length becomes smaller than the characteristic linear scale of the system, which happens at finite temperatures. However, in Ref.~\cite{Starkov_2026}, it was demonstrated analytically for the case of a straight channel, that the correction to the conductivity at small temperatures is always positive, if the even-odd effect is taken into account, and grows as $\propto T^2$. The same superballistic effect has been demonstrated numerically in Ref.~\cite{Estrada-Alvarez_2025} for the general crenellated channels, which has finally provided a proper explanation for a number of transport experiments~\cite{Sarypov_2025,Keser_2021,KrishnaKumar_2017,Renard_2008,Ginzburg_2021,Estrada-Alvarez_2025b,Ginzburg_2023}.

Recently, there has been an experiment on magnetotransport in the Corbino geometry, where the scaling of kinematic viscosity $\nu\propto 1/T$ was reported, in stark contrast to the Fermi liquid prediction $\nu\propto 1/T^2$. This anomalous scaling was attributed to the tomographic flow regime based on the results of Ref.~\cite{Kryhin_2025}. The crucial assumption of the scaling analysis was the existence of a very large number of the long-lived harmonics, which enabled the authors to treat the mode number as a continuous variable. Yet, for the typical temperatures in the experiment~$T/E_F\sim 10^{-2}-10^{-1}$, the number of the long-lived harmonics is estimated to be $<5$, which sheds doubt on the applicability of the results of Ref.~\cite{Kryhin_2025} and raises the question whether at all the even-odd effect is responsible for the anomalous scaling.

As we have discussed, the effects related to the long-lived odd harmonics may be easier to spot in the ballistic and at the onset of the tomographic regimes. With that in mind, in our work we numerically solve the linearized Boltzmann equation for the magnetotransport in the Corbino disk geometry across all the flow regimes to identify the key signature that can be attributed to the even-odd effect. 
We focus on the resistance sensitivity with respect to the magnetic field $\alpha = \partial R/\partial(B^2)$ --- the quantity that has been used in the experiment~\cite{Zeng_2024} to extract the relevant parameters. As we demonstrate, the presence of the long-lived odd harmonics leads to the enhancement of $\alpha$ at $B\rightarrow0$. Moreover, this enhancement is most prominent at the onset of the tomographic regime, $v_F \tau_{e}^{(\mathrm{ee})}\sim r_b$, where $r_b$ is the outer radius of the Corbino disk.

The paper is structured as follows. In Section~\ref{sec:model}, we discuss the theoretical model for the flow in the Corbino disk based on the linearized Boltzmann equation. In Section~\ref{sec:results}, we present the results of our numerical computations, which we discuss in Section~\ref{sec:discussion}. The Appendix~\ref{details-integral-eq} describes how the method of characteristics can be extended to the case of multiple long-lived odd harmonics to recast the Boltzmann equation as a system of integral equations. This approach was used to test the direct numerical method based on the discretization of the Boltzmann equation. The latter we explain in Appendix~\ref{details-discretization}.

{\it Note:} while we were finishing the manuscript, a work appeared~\cite{Ben-Shachar_2026}, where the same enhancement of the magnetotransport coefficient $\alpha$ at small magnetic fields is predicted for the Corbino geometry. The approach in that work is applicable only when the size of the boundary layers $v_F\sqrt{\tau_e^{(\mathrm{ee})}\tau_o^{(\mathrm{ee})}}$ becomes much smaller than the typical size of the disk $r_b$, which restricts it to higher temperatures. On the contrary, we study the problem across all regimes, which includes the ballistic and the crossover to the hydrodynamic one.

\section{Model}
\label{sec:model}

\begin{figure}[t]
    \includegraphics[width=0.8\columnwidth]{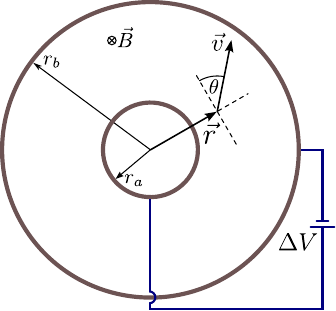}
    \caption{Schematic representation of a Corbino disk.}
    \label{Corbino-scheme}
\end{figure}

We consider a flat sheet of conductive 2D material between two massive leads in the form of rings of radii $r_a$ and $r_b$, in the presence of a magnetic field $\vec B$ normal to the material plane (see Fig.~\ref{Corbino-scheme}).
A voltage difference $\Delta V = \Phi_a-\Phi_b$ is applied between the leads, where $\Phi_{a,b}$ are electrostatic potentials of the leads. We are going to assume that $\Delta V$ is sufficiently small, so the electron distribution $f=f_0+\delta f$ within the 2D material can be described by a linearized Boltzmann equation:
\begin{equation}
    \vec v \cdot\vec\nabla_r \delta f + e\left(\frac{\vec v}{c}\times B\right)\cdot \vec\nabla_p \delta f + I_c{\{\delta f\}} = \left(-\frac{\partial f_0}{\partial\epsilon}\right)e\vec E\cdot\vec v,\label{lboltzmann}
\end{equation}
where $I_c{\{\delta f\}}$ is the linearized collision operator and $f_0$ is the equilibrium Fermi-Dirac distribution at temperature $T$. The electric field $\vec E= -\vec\nabla_r \Phi$ is expressed in the usual manner via the electrostatic potential $\Phi$, which depends only on the distance to the center of the Corbino disk $r$ due to the axial symmetry of the problem.
In principle, the voltage difference between the leads results in the redistribution of charges, and to determine $\Phi(r)$, we would need to solve Eq.~\eqref{lboltzmann} together with the Poisson equation, that relates $\Phi$ to the excess charge density
\begin{equation}
    \delta \rho = g_s g_v e\int \frac{d^2 \vec k}{(2\pi)^2} \delta f,
\end{equation}
where $g_s$ and $g_v$ are the spin and valley degeneracy factors respectively.
Once $\delta f$ is known, we can find the normal and tangential components of the current density:
\begin{align}
j_n(r) &= g_s g_v e\int \frac{d^2\vec k}{(2\pi)^2} \delta f v \sin{\theta},\\
j_t(r) &= g_s g_v e\int \frac{d^2\vec k}{(2\pi)^2} \delta f v \cos{\theta}.
\end{align}
Here, $\theta$ is the angle between the velocity vector $\vec v$ and the tangential direction to the radius-vector $\vec r$ (see Fig.~\ref{Corbino-scheme}).
The total normal current is given by
\begin{equation}
    J_n = 2\pi r j_n(r)\equiv \mathrm{const}
\end{equation}
and is constant inside the disk due to the conservation of charge.
The conductance of the Corbino disk is then
\begin{equation}
    G = J_n/\Delta V.
\end{equation}

In the linearized regime, the resulting conductance depends only on the difference $\Delta V = \Phi_a-\Phi_b$ and not on the values $\Phi_{a,b}$ themselves. For the sake of concreteness, we have chosen the lead potentials in a symmetric fashion,
\begin{equation}
    \Phi_a = -\Phi_b = \frac{\Delta V}{2}.
\end{equation}

To simplify Eq.~\eqref{lboltzmann}, it is convenient to use the ansatz
\begin{equation}
    \delta f = \left(-\frac{\partial f_0}{\partial \epsilon}\right) \left[\eta -e\Phi\right]
\end{equation}
In the assumption of the degenerate electron gas $T\ll E_F$, $(-\partial f_0/\partial\epsilon)$ is highly peaked at the Fermi energy, so we can neglect the energy dependence of $\eta$ and pin all velocities to the Fermi level.
The resulting equation takes the form
\begin{multline}
\sin{\theta}\frac{\partial \eta}{\partial r} + \left(\frac{\cos{\theta}}{r} - R_L^{-1}\right)\frac{\partial \eta}{\partial \theta} +\\+ v_F^{-1}\int_{0}^{2\pi}I(\theta-\theta^\prime) \eta(\theta^\prime) = 0,\label{lboltzmann-simplified}
\end{multline}
where we have defined the Larmor radius $R_L = \hbar k_F c/(|e|B)$ using the effective electron mass $m^*$.

Several comments are in order. First, due to the axial symmetry, $\eta$ depends only on the relative angle between $\vec r$ and $\vec v$ which is directly expressed via $\theta$. As a result, $\eta\equiv \eta(r,\theta)$.
Secondly, after we pin the physics to the Fermi level, the linearized collision operator becomes an integral operator describing the angular diffusion on the Fermi surface. The action of this integral operator on $\eta$ is captured by the third term in Eq.~\eqref{lboltzmann-simplified}.
If we assume the Fermi surface to be circular, we can expand both the integral kernel $I(\theta-\theta^\prime)$ and $\eta$ in circular harmonics:
\begin{align}
    \eta(r,\theta) &= \eta_0(r) + \sum_{m=1}^{+\infty} \left[\eta_m^{(c)}(r)\cos{m\theta} + \eta_m^{(s)}(r)\sin{m\theta}\right],\\
    I(\theta-\theta^\prime) &= \frac{1}{2\pi \tau_0} + \sum_{m=1}^{+\infty} \frac{\cos{m(\theta-\theta^\prime)}}{\pi \tau_m}.\label{I-expand}
\end{align}
These circular harmonics are the eigenmodes of the collision operator, and the inverse scattering times $\tau_m^{-1}$ are the corresponding eigenvalues. Due to charge conservation, $\tau_0=+\infty$ and the first term in Eq.~\eqref{I-expand} simply falls out.
Throughout the paper, we will also refer to the inverse scattering times as scattering rates.

Note that the electric field has been removed from Eq.~\eqref{lboltzmann-simplified} completely. Since $e\Phi(r)$ does not depend on the angle $\theta$, this shift only affects the zeroth harmonic $\eta_0(r)$, and we can use $\delta \tilde f = (-\partial f_0/\partial\epsilon)\eta$ to compute the current density and other observables attributed to the higher-order angular harmonics. This way, there is no need to solve the Poisson equation, which simplifies things considerably. The zeroth harmonic itself corresponds physically to the non-equilibrium part of the electrochemical potential, 
\begin{equation}
    \eta_0(r) = \bar\mu(r) = \delta\mu(r)+e\Phi(r).\label{electrochemical-potential}
\end{equation}

The fact that the voltage difference is applied to the leads is encoded now in the boundary conditions. We are going to model the latter in the spirit of Fuchs~\cite{Fuchs_1938}: we will assume that electrons have an angular-dependent probability $r_\theta$ to scatter specularly at the interface with the lead. Correspondingly, $(1-r_\theta)$ is the probability that the electrons cross the interface. Thus, we obtain, for example, for the inner lead
\begin{multline}
    \eta(r_a+0,\theta)-e\Phi(r_a+0) =\\ (1-r_\theta)\left[\eta(r_a-0, \theta)-e\Phi(r_a-0)\right] + \\ +r_\theta\left[\eta(r_a+0, 2\pi-\theta)-e\Phi(r_a+0)\right],\label{inner-boundary-raw}
\end{multline}
where the angles $0\leqslant\theta\leqslant \pi$ correspond to the electrons moving outwards from the lead. The quantities at $r_a-0$ correspond to the lead side, while the ones at $r_a+0$ describe the electrons in the 2D material.
The electric potential is taken to be continuous at the interface~\footnote{A potential jump would imply the presence of a singular charge layer at the interface. Such a singular structure should not appear at the kinetic level of description, where the boundary region is resolved explicitly.}: $\Phi(r_a-0)=\Phi(r_a+0)=\Phi_a$. Since it does not depend on $\theta$, we can drop $e\Phi(r)$ from Eq.~\eqref{inner-boundary-raw} entirely. Assuming that the lead is massive, we can neglect the change in the distribution function there, so that $\eta(r_a-0,\theta)=e\Phi(r_a)$. This casts Eq.~\eqref{inner-boundary-raw} as
\begin{equation}
    \eta(r_a,\theta) = (1-r_\theta)e\Phi_a + r_\theta \eta(r_a,2\pi-\theta).\label{inner-boundary}
\end{equation}
Similar analysis leads to the boundary condition at the interface with the outer lead:
\begin{equation}
    \eta(r_b, 2\pi-\theta) = (1-r_\theta)e\Phi_b + r_\theta \eta(r_b,\theta),\label{outer-boundary}
\end{equation}
where again $0\leqslant\theta\leqslant \pi$.

We consider two types of scattering processes that contribute to the collision operator.
The first type corresponds to the processes that result in the relaxation of momentum, such as impurity scattering or electron-phonon scattering. We model their effect by a mode-independent scattering time $\tau_m^{(\mathrm{mr})} = \tau^{(\mathrm{mr})}$ for $m\geqslant 1$. As we mentioned, the scattering time for the zeroth harmonic is infinite due to charge conservation.
The second type corresponds to the electron-electron collisions that conserve the total momentum~\footnote{The Umklapp collisions can still relax the momentum, however, their contribution is small unless the Fermi surface is large.}, resulting in the scattering times $\tau_m^{(\mathrm{ee})}$. Notice that both $\tau_0^{(\mathrm{ee})}$ and $\tau_1^{(\mathrm{ee})}$ are infinite due to the charge and momentum conservation respectively.
The total scattering rate is given by Matthiessen’s rule,
\begin{equation}
\frac{1}{\tau_m} = \frac{1}{\tau_m^{(\mathrm{mr})}}+\frac{1}{\tau_{m}^{(\mathrm{ee})}}.\label{eq:mathiessen}
\end{equation}

As we have discussed in the introduction, the scattering rates for even harmonics only weakly depend on the mode number and scale as the square of the temperature (see Eq.~\eqref{base-even-rate})
At the same time, for the odd scattering rates at small $k$ one obtains:
\begin{equation}
    1/\tau_{2k+1}^{(\mathrm{ee})} = (2k+1)^4/\tau_{o}^{(\mathrm{ee})},\label{odd-rate-small}
\end{equation}
where $1/\tau_o^{(\mathrm{ee})}$ is given by Eq.~\eqref{base-odd-rate}.
As we see, it contains another $(T/E_F)^2\ll1$ prefactor, while the odd scattering rates quickly grow with the harmonic number. For large $k$, they saturate at the level of even relaxation rates~\cite{Ledwith_2019a, Hofmann_2023, Nilsson_2025}, which can be modelled by~\cite{Kryhin_2025}
\begin{equation}
    1/\tau_{2k+1}^{(\mathrm{ee})} = \frac{1/\tau_{e}^{(\mathrm{ee})}}{1 + \frac{\tau_o^{(\mathrm{ee})}}{\tau_e^{(\mathrm{ee})}}\frac{1}{(2k+1)^4}}.\label{eq:odd-interpolation}
\end{equation}
This agrees with Eq.~\eqref{odd-rate-small} at small $k$.

The number $k^\prime$ of the ``long-lived'' odd harmonics with the reduced scattering rates is sensitive to the temperature. We can estimate $k^\prime$ as
\begin{equation}
    k^\prime\sim \frac{1}{2} \sqrt{\frac{E_F}{T}}.
\end{equation}
In Table~\ref{table:k-dependence}, we display the estimates for different $T/E_F$ ratios.
\begin{table}[h]
\caption{Estimated number of long-lived odd harmonics for different temperatures.}
\label{table:k-dependence}
\begin{tabular}{|c|c|c|c|c|}\hline
$T/E_F$ & $10^{-4}$ & $10^{-3}$ & $10^{-2}$ & $10^{-1}$ \\ \hline
$k^\prime$ & 50 & 16 & 5 & 2 \\ \hline
\end{tabular}
\end{table}
As we see, for moderately small temperatures $T/E_F\sim 10^{-3}-10^{-2}$, the number of the long-lived harmonics is only of the order of $10$.

To analyze the different flow regimes, we shall introduce the dimensionless scattering rates $\gamma = r_b/(v_F\tau)$.
The constants of proportionality in Eqs.~\eqref{base-even-rate} and~\eqref{base-odd-rate} depend on the dimensionless interaction strength $r_s=1/(\sqrt{2}k_F l_\mathrm{TF})$, where $l_\mathrm{TF}$ is the Thomas-Fermi screening length, and the relative ratio of the odd to even scattering rates can be tuned by changing $r_s$~\cite{Nilsson_2025}. To keep the discussion less dependent on all the concrete details, it is convenient to express the temperature through the even relaxation rate using Eq.~\eqref{base-even-rate}
\begin{equation}
T/E_F \propto \sqrt{\gamma_e^{(\mathrm{ee})}}.
\end{equation}
This way we can cast the dependence of the odd scattering rates on temperature as~\cite{Starkov_2026}
\begin{equation}
    \gamma_{o}^{(\mathrm{ee})}/\gamma_{e}^{(\mathrm{ee})} = a^{-1} \gamma_e^{(\mathrm{ee})},\label{a-def}
\end{equation}
where $a$ is the phenomenological parameter that governs how quickly odd scattering rates catch up with the even scattering rate as the latter grows. 
Combining Eqs.~\eqref{eq:mathiessen}, \eqref{base-even-rate}, \eqref{eq:odd-interpolation} and~\eqref{a-def}, we can then write the total dimensionless scattering rates as
\begin{equation}
    \gamma_0 = 0,\qquad \gamma_1 = \gamma^{(\mathrm{mr})},
\end{equation}
\begin{equation}
    \gamma_{2k} \equiv \gamma^{(\mathrm{mr})} + \gamma_e^{(\mathrm{ee})},\qquad k\geqslant1,
\end{equation}
\begin{equation}
    \gamma_{2k+1} = \gamma^{(\mathrm{mr})} + \frac{(2k+1)^4 a^{-1}\gamma_{e}^{(\mathrm{ee})}}{1 + (2k+1)^4 a^{-1}\gamma_{e}^{(\mathrm{ee})}}\gamma_e^{(\mathrm{ee})}, \qquad k\geqslant 1.\label{eq:scattering-rates-full}
\end{equation}
Making this parameter $a$ larger effectively increases the number of the long-lived harmonics.

One can also express the estimate for the typical number of the long-lived harmonics in the tomographic regime through $a$. In this regime, $\sqrt{\gamma_e^{(\mathrm{ee})}\gamma_o^{(\mathrm{ee})}} \sim 1$, which implies $\gamma_e^{(\mathrm{ee})}\sim a^{1/3}$ and
\begin{equation}
    k^\prime \sim \frac{1}{2}\sqrt{\frac{E_F}{T}} = \frac{1}{2}\sqrt{\frac{\gamma_e^{(\mathrm{ee})}}{\gamma_o^{(\mathrm{ee})}}}\sim \frac{a^{1/3}}{2}.
\end{equation}

Assuming that the constants of proportionality in Eqs.~\eqref{base-even-rate} and~\eqref{base-odd-rate} are of the order $O(1)$, we can roughly estimate
\begin{equation}
    a\sim \frac{E_F r_b}{hv_F}\sim k_F r_b.\label{a-estimate}
\end{equation}
More accurate calculations provided in the supplementary~\cite{supplement} suggest that the values of $a$ in the experiment~\cite{Zeng_2024} were device-dependent and varied in the range $1000-5000$.

\section{Numerical Results}\label{sec:results}

\begin{figure}[t]
\centering
\includegraphics[width=0.7\columnwidth]{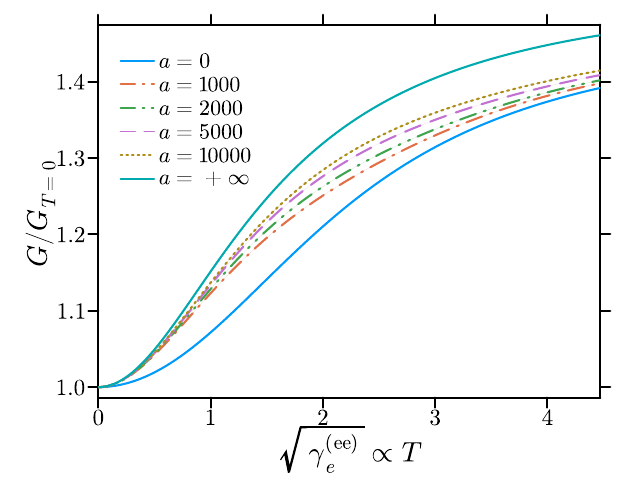}
\caption{Normalized total conductance $G/G_{T=0}$ of the Corbino disk as a function of $\sqrt{\gamma_e^{(\mathrm{ee})}}\propto T$ for different values of the phenomenological parameter $a$ (see Eqs.~\eqref{a-def} and~\eqref{a-estimate} for the discussion of its physical meaning). The case $a=0$ corresponds to the Callaway's dual relaxation time approximation, while the case $a=+\infty$ describes the model with an infinite number of the long-lived harmonics.}
\label{fig:temperature-dependence}
\end{figure}

Unlike the case of a straight channel, here the analytical solution is possible only when the overall scattering length~$l$ becomes much smaller than the radii $r_{a,b}$ of the Corbino disk, in which case the transport is described by the Stokes-Ohm equation~\cite{Shavit_2019, Levchenko_2020, Zeng_2024, Weiss_1954, Blood_1971, Wieder_1969}. For the straight channel, we can ignore the zeroth harmonic of the distribution function, and then the method of characteristics provides us with the exact solution in the ballistic case~\cite{Starkov_2026}. In the Corbino geometry, however, this is not possible due to the nontrivial redistribution of the charges.

In order to analyze the flows across the different regimes, the only option available is resorting to numerics. Here, one possible approach is to extend the method of characteristics to the case when there are multiple long-lived harmonics. As a result, the linearized Boltzmann equation is recast as a system of integral equations that can be solved numerically. We explain the details of this approach in Appendix~\ref{details-integral-eq}. The downside of this method is that the number of components of the kernel grows quadratically with the number of long-lived harmonics, which makes it impractical to model the ballistic regime and the onset of the tomographic one. On the other hand, the kernels only depend on the dimensionless scattering rate $\gamma$ at large mode numbers. Thus, having computed the kernels once, we can play with the scattering rates for the long-lived harmonics however we want.

Another approach is based on the direct discretization of the linearized Boltzmann equation. We discuss this method in detail in Appendix~\ref{details-discretization}. Essentially, this is an adaptation of the method used by us in Ref.~\cite{Starkov_2026} to the case of the Corbino geometry.

In the paper, we present the results obtained with the direct discretization approach, because it is computationally more efficient. However, we did implement both methods and compared the obtained solutions against each other for a selected set of parameters to make sure that the results agree.

Throughout all the computations, we kept $r_b/r_a=4$, since this ratio is the one used in the experiment of Ref.~\cite{Zeng_2024}. In addition to that we used a fixed $\gamma^{(\mathrm{mr})}=0.2$ and assumed fully diffusive boundary conditions $r_\theta\equiv 0$ at the interfaces with the leads.

\subsection{Temperature dependence in the absence of magnetic field}

First, we consider changing the even scattering rate $\gamma_e^{(\mathrm{ee})}$ at zero magnetic field. In Fig.~\ref{fig:temperature-dependence}, we present the total conductance of the Corbino disk $G/G_{T=0}$ normalized by its value in the absence of electron-electron collisions as a function of $\sqrt{\gamma_e^{(\mathrm{ee})}}\propto T/E_F$. The results are presented for different values of the phenomenological parameter $a$ including $a=0,+\infty$. The limit $a\rightarrow 0$ in Eq.~\eqref{eq:scattering-rates-full} results in the Callaway's dual relaxation time approximation for the collision integral~\cite{Callaway_1959}: $\gamma_1 = \gamma^{(\mathrm{mr})}$, $\gamma_{m\geqslant2} \equiv \gamma^{(\mathrm{mr})}+\gamma_e^{(\mathrm{ee})}$. This is the standard model that does not take into account the long-lived harmonics. The opposite limit $a\rightarrow +\infty$ corresponds to a purely theoretical model where the electron-electron collisions are assumed not to relax any of the odd harmonics: $\gamma_{2k-1}=\gamma^{(\mathrm{mr})}$, $\gamma_{2k} = \gamma^{(\mathrm{mr})}+\gamma_e^{(\mathrm{ee})}$ for $k=1,2,3,\dotsc,+\infty$.

As we see, in the Corbino geometry, the conductance grows with the temperature even in the dual relaxation time approximation. This is in stark contrast to the case of the straight channel, where we expect the conductance to decrease at small temperatures if we neglect the long-lived harmonics~\cite{Starkov_2026}. When the value of the parameter $a$ is increased, the number of the long-lived harmonics effectively grows. As we observe in Fig.~\ref{fig:temperature-dependence}, it speeds up the increase in the conductance with $\sqrt{\gamma_e^{(\mathrm{ee})}}$.

\subsection{Magnetic-field dependence}

\begin{figure}[t]
\centering
\includegraphics[width=\columnwidth]{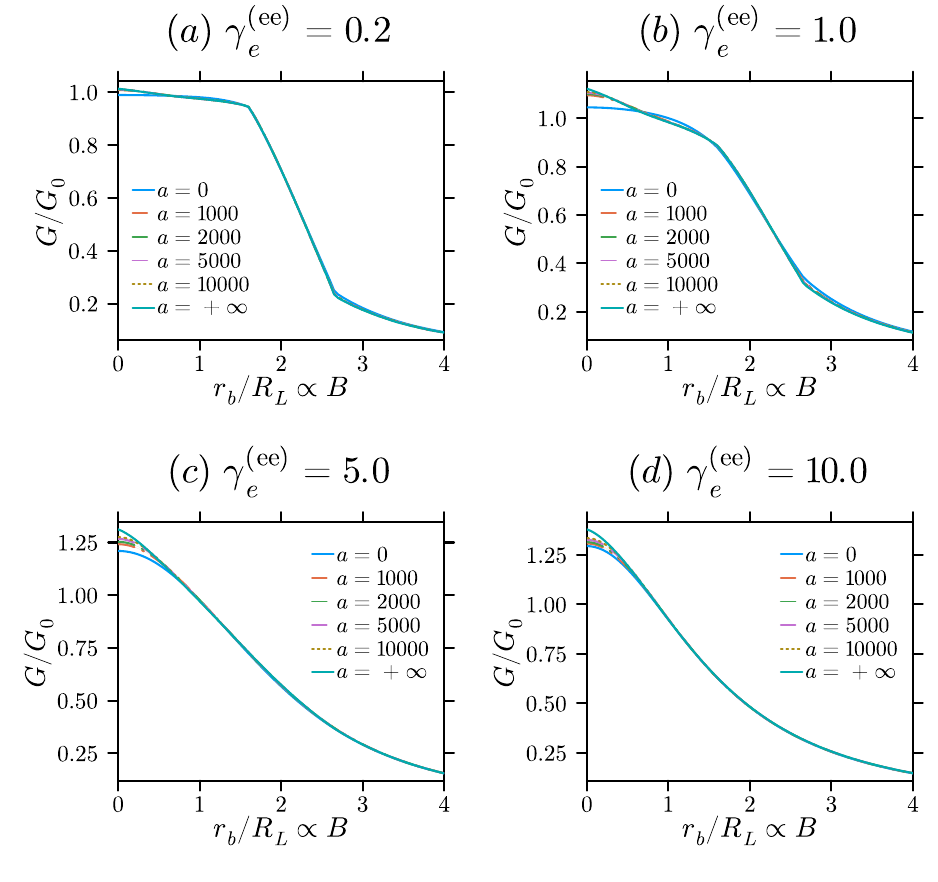}
\caption{Normalized total conductance $G/G_0$ of the Corbino disk as a function of the inverse Larmor radius $r_b/R_L\propto B$ for different values of the phenomenological parameter $a$ and the even electron-electron scattering rate $\gamma_{e}^{(\mathrm{ee})}$. The case $a=0$ corresponds to the Callaway's dual relaxation time approximation, while the case $a=+\infty$ describes the model with an infinite number of the long-lived harmonics. See the main text for the definition of $G_0$.}
\label{fig:magfield-dependence}
\end{figure}

The main objective of the paper is to analyze the influence of the long-lived odd harmonics on the magnetotransport. In Fig.~\ref{fig:magfield-dependence}, we display the dependence of the normalized total conductance $G/G_0$ on the inverse Larmor radius $r_b/R_L$, which is linearly proportional to the magnetic field, for different values of the even electron-electron scattering rate $\gamma_e^{(\mathrm{ee})}$.
The normalization constant is
\begin{equation}
G_0=\nu_F e^2v_Fr_b/2,
\end{equation}
where
\begin{equation}
    \nu_F = \frac{g_sg_v m^*}{2\pi \hbar^2}
\end{equation}
is the density of states at the Fermi level including the spin (and/or valley) degeneracy.
Again, we compare the curves obtained for different values of the phenomenological parameter $a$.

Overall, the difference between the case with a nonzero number of the long-lived harmonics ($a>0$) and the dual relaxation time approximation $(a=0)$ quickly vanishes as the magnetic field increases. This is in agreement with the prediction of Ref.~\cite{Rostami_2025} that the magnetic field should destroy the tomographic flow regime. To understand this result qualitatively, it is convenient to switch to the basis of complex angular harmonics:
\begin{equation}
    \eta(r, \theta) = \sum_{m=-\infty}^{+\infty} \eta_m(r)e^{im\theta}.
\end{equation}
In this representation, the effects of the magnetic field can be taken into account by renormalizing the scattering rates:
\begin{equation}
    \gamma_m \rightarrow \tilde\gamma_m = \gamma_m - im\frac{r_b}{R_L},\label{eq:gamma-renormalization}
\end{equation}
where it is assumed $\gamma_{-m}=\gamma_m$.
Starting with some mode number, the magnetic field completely dominates the collision processes. Therefore, when the magnetic field is increased, this cutoff mode number goes down, masking the effects of the higher-order long-lived odd harmonics.

\begin{figure}[t]
\centering
\includegraphics[width=\columnwidth]{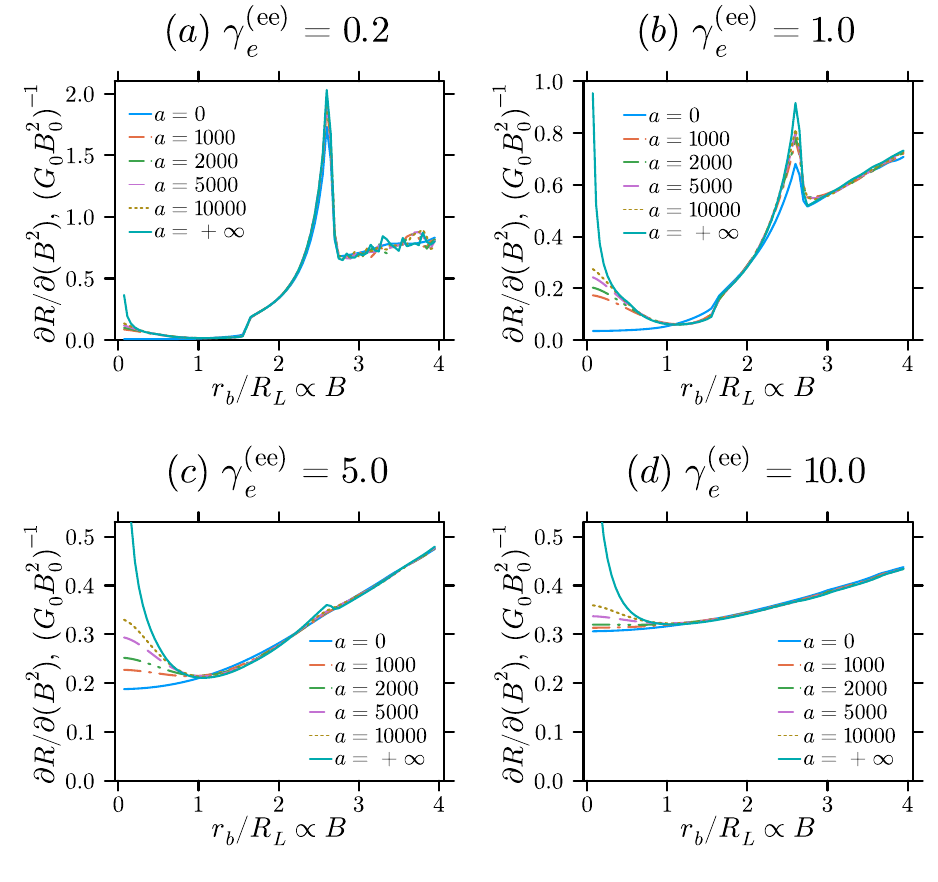}
\caption{Sensitivity $\partial R/\partial(B^2)$ of the resistance $R=1/G$ to the square of the magnetic field as a function of the inverse Larmor radius $r_b/R_L\propto B$ for different values of the phenomenological parameter $a$ and the even electron-electron scattering rate $\gamma_e^{(\mathrm{ee})}$. The case $a=0$ corresponds to the Callaway's dual relaxation time approximation, while the case $a=+\infty$ describes the model with an infinite number of the long-lived harmonics. See the main text for the definition of $G_0$ and $B_0$.}
\label{fig:magfield-sensitivity}
\end{figure}

To draw direct parallels to the analysis of Ref.~\cite{Zeng_2024}, we have also computed the sensitivity $\alpha = \partial R/\partial (B^2)$ of the resistance $R=1/G$ to the square of the magnetic field and display the results in Fig.~\ref{fig:magfield-sensitivity}. This is precisely the quantity that was used in Ref.~\cite{Zeng_2024} to extract the parameters of the flow. The results are shown in units of $(G_0B_0^2)^{-1}$, where
\begin{equation}
    B_0 = \frac{\hbar k_F c}{|e|r_b}
\end{equation}
is the value of the magnetic field, for which the Larmor radius is exactly $r_b$.

The main effect of the long-lived odd harmonics is the enhancement of the sensitivity $\alpha$ at zero magnetic field, which is absent for the dual relaxation time approximation ($a=0$).
It is especially large for a model with an infinite number of the long-lived harmonics ($a=+\infty$) in the regime where $\gamma_e^{(\mathrm{ee})}\gtrsim 1$. In this case it looks as if $\alpha$ diverges in the limit $B \rightarrow 0$. For realistic values of $a$, the enhancement seems to be most prominent at the onset of the tomographic flow regime when $\gamma_{e}^{(\mathrm{ee})}\sim 1$ and gradually disappears when $\gamma_{e}^{(\mathrm{ee})}$ is increased deeper into the hydrodynamic regime.

In addition to that, in the ballistic regime and at the onset of the tomographic one, $\gamma_e^{(\mathrm{ee})}\lesssim 1$, the sensitivity $\alpha$ exhibits a sharp peak at $r_b/R_L\approx2.67$, whose height is sensitive to the presence of the long-lived harmonics. This feature corresponds to the value of the magnetic field, for which a circular electron trajectory fits exactly between two leads, {\it i.e.\/}, $2R_L=r_b-r_a$.

We have also checked how the normalized conductance and resistance sensitivity behave at larger values of the momentum-relaxing scattering rate $\gamma^{(\mathrm{mr})}=0.5$ and $\gamma^{(\mathrm{mr})}=1.0$.
The results seem to be qualitatively the same, although it looks like increasing $\gamma^{(\mathrm{mr})}$ makes the peak in the sensitivity less prominent~\cite{supplement}.

\subsection{Electrochemical-potential and tangential-current profiles at small magnetic field}

\begin{figure}[t]
\centering
\includegraphics[width=\columnwidth]{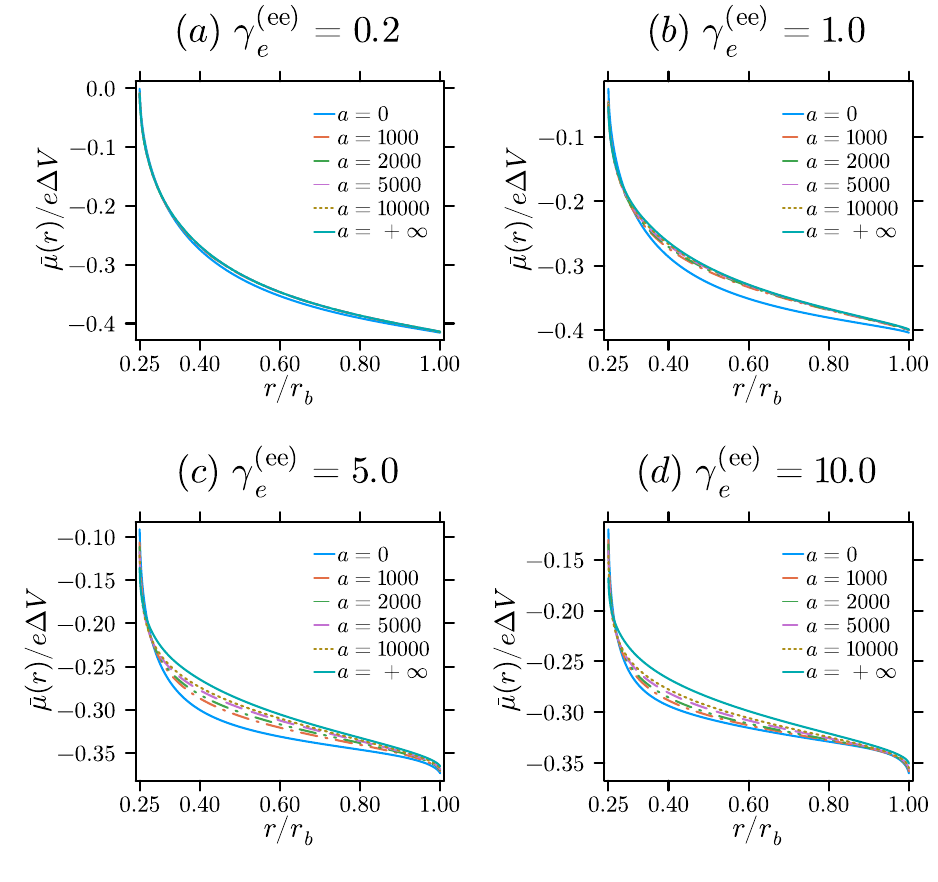}
\caption{Normalized electrochemical potential $\bar\mu(r)/e\Delta V = (e\Phi(r)+\delta\mu(r))/e\Delta V$ inside the Corbino disk as a function of the distance $r$ to the disk center for different values of the phenomenological parameter $a$ and the even electron-electron scattering rate $\gamma_e^{(\mathrm{ee})}$. The case $a=0$ corresponds to the Callaway's dual relaxation time approximation, while the case $a=+\infty$ describes the model with an infinite number of the long-lived harmonics.}
\label{fig:potential-profiles}
\end{figure}

\begin{figure}[t]
\centering
\includegraphics[width=\columnwidth]{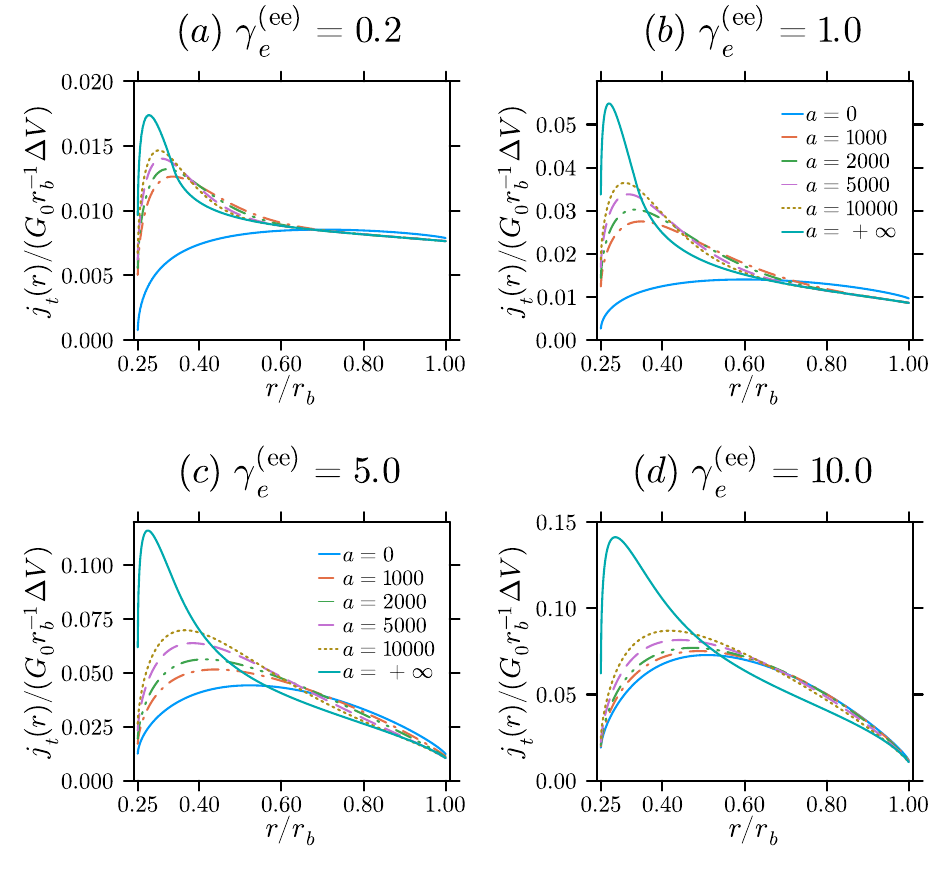}
\caption{Normalized tangential current density $j_t(r)/(G_0r_b^{-1}\Delta V)$ inside the Corbino disk as a function of the distance $r$ to the disk center for different values of the phenomenological parameter $a$ and the even electron-electron scattering rate $\gamma_e^{(\mathrm{ee})}$. The case $a=0$ corresponds to the Callaway's dual relaxation time approximation, while the case $a=+\infty$ describes the model with an infinite number of the long-lived harmonics.}
\label{fig:tang-current-profiles}
\end{figure}

It is also interesting to directly visualize the influence of the long-lived odd harmonics on the distribution function within the disk. To this end, we plot the normalized electrochemical potential $\bar\mu(r)/e\Delta V = (e\Phi(r)+\delta\mu(r))/e\Delta V$ (see the discussion around Eq.~\eqref{electrochemical-potential}) and the normalized tangential current $j_t(r)/(G_0r_b^{-1}\Delta V)$ at a fixed value of $r_b/R_L=0.1$ in Figs.~\ref{fig:potential-profiles} and~\ref{fig:tang-current-profiles} respectively.

The normalized electrochemical potential of the leads is
\begin{equation}
    \frac{\bar\mu_{a,b}}{e\Delta V} = \frac{\Phi_{a,b}}{\Delta V} = \pm 1/2
\end{equation}
As we can see in Fig.~\ref{fig:potential-profiles}, the normalized electrochemical potential does not reach the values $\pm 1/2$ when approaching the leads from within the Corbino disk. The electrochemical potential experiences jumps at the interfaces with the leads that can be attributed to the interface resistance.
Another interesting observation is that the tangential current does not vanish at the interfaces with the leads even for $\gamma_e^{(\mathrm{ee})}=10.0$, which implies a finite value of the slip length.

Overall, the long-lived odd harmonics influence both the distribution of the electrochemical potential and the distribution of the tangential current density within the Corbino disk, however, the effect is most prominent in the latter case, where it leads to an elevated tangential current density close to the inner lead (see Fig.~\ref{fig:tang-current-profiles}).

\section{Discussion}\label{sec:discussion}

In this work, we analyzed the electron flows in the two-dimensional Corbino disk geometry across all the flow regimes and demonstrated that the main qualitative signature of the even-odd effect is the enhancement of the resistance sensitivity $\alpha =\partial R/\partial(B^2)$ at magnetic fields around $0$. The enhancement is most prominent at the onset of the tomographic regime $\gamma_e^{(\mathrm{ee})} = r_b/(v_F\tau_e^{(\mathrm{ee})})\sim 1$. When the even scattering rate is increased (by making temperature larger) into the hydrodynamic regime, this enhancement gradually disappears.
Interestingly enough, if we consider a purely theoretical model where an infinite number of the long-lived harmonics is present, $\alpha$ appears to diverge at $B\rightarrow 0$ and the enhancement survives deep into the hydrodynamic regime.

The analysis in the experimental paper~\cite{Zeng_2024} assumed that $\alpha=\partial R/\partial(B^2)$ experiences one plateau at $B=0$ and another one at larger values of the magnetic field. The two values of $\alpha$ extracted at small and large magnetic fields were then used to obtain the bulk resistivity due to momentum-relaxing processes and the kinematic viscosity.
Without access to the raw experimental data, it is hard to argue whether the enhancement has not been observed at all or it was simply not accounted for.
In the latter case, the enhancement due to the even-odd harmonics could have influenced the value of $\alpha$ extracted at small magnetic fields and the value of kinematic viscosity computed from it. As such, it could explain the anomalous scaling of the kinematic viscosity with temperature $\nu\propto 1/T$, as it is suggested in Ref.~\cite{Ben-Shachar_2026}.

This anomalous scaling was reported in the temperature range $50-200\ \mathrm{K}$.
According to our estimates~\cite{supplement}, the typical $\gamma_e^{(\mathrm{ee})}$ for the samples with bilayer graphene was of the order of several tens.
In this range, the enhancement becomes relatively small and is easy to overlook, which implies that the observed $\nu\propto 1/T$ scaling might have nothing to do with the even-odd effect.
In the case of monolayer graphene, some samples could have $\gamma_{e}^{(\mathrm{ee})}$ closer to $1$ at the lower end of that temperature range, however the typical $\gamma_{e}^{(\mathrm{ee})}$ was around $5$. Although the enhancement plays a larger role for these values of $\gamma_{e}^{(\mathrm{ee})}$, it does not constitute the dominating contribution to the value of $\alpha$ at small magnetic fields. Therefore, it does not seem clear whether the enhancement can completely explain the discrepancy between the observed $\nu\propto 1/T$ and expected from the Fermi-liquid theory $\nu\propto 1/T^2$ scalings of the kinematic viscosity in this case.

In either case, the experimental observation of the predicted enhancement of the magnetotransport coefficient $\alpha$ should be a clear demonstration of the even-odd effect. As our analysis suggests, increasing the phenomenological parameter $a\sim k_F r_b$ is crucial to make the enhancement more visible. This implies that having larger charge density $n$ and samples of bigger size $r_b$ is beneficial in this regard. One should bear in mind, however, that the even-odd effect is sensitive to the presence of the inversion symmetry $\varepsilon_{\vec k} = \varepsilon_{-\vec k}$, which can be destroyed due to the trigonal warping which plays a more prominent role at larger charge doping.

\section{Acknowledgment}
I would like to thank Prof. Dr. Björn Trauzettel for the fruitful discussions.
I also acknowledge support by DFG-SFB 1170 (Project-ID: 258499086) and EXC2147 ctd.qmat (Project-ID: 390858490).

\appendix

\section{Integral form of the linearized Boltzmann equation\label{details-integral-eq}}
\subsection{Theory}

\begin{figure}[t]
\centering
\includegraphics{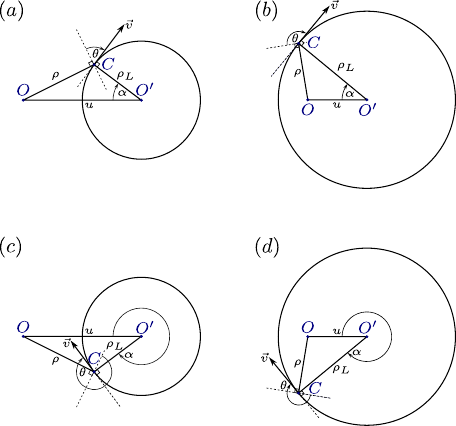}
\caption{Triangle formed by the point $C$ where the distribution function is analyzed, center $O$ of the Corbino disk and center $O^\prime$ of the electron trajectory going through the point $C$, for the case when $(a,c)$ point $O$ lies outside the circular trajectory; $(b,d)$ point $O$ lies inside the circular trajectory. Geometrical configurations of panels $(c,d)$ are obtained from those of the panels $(a,b)$ by mirror reflecting the point $C$ with respect to $OO^\prime$. As a result, the angles satisfy $\theta_{c,d}=2\pi-\theta_{a,b}$, $\alpha_{c,d}=2\pi-\alpha_{a,b}$, where the indices refer to the different panels.}
\label{trajectory-sketch}
\end{figure}

\begin{figure}[t]
\centering
\includegraphics{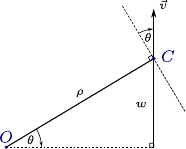}
\caption{Geometrical sketch of an electron trajectory in the case of zero magnetic field.}
\label{zero-field-sketch}
\end{figure}

In this Appendix, we are going to show how the method of characteristics can be used to transform the integro-differential equation~\eqref{lboltzmann-simplified} into a system of purely integral equations even when the long-lived odd harmonics are taken into account. The idea of our approach is similar to the one used in Refs.~\cite{Raichev_2020, Raichev_2022, Raichev_2022a, Raichev_2023}, however we want to emphasize that, to the best of our knowledge, this method has not been applied to the case where multiple (more than two) different relaxation times appear in the expansion~\eqref{I-expand}.

We are going to assume that only a finite number of the long-lived harmonics is present. This assumption is satisfied for moderately small temperatures, as we have discussed in Section~\ref{sec:model}. It means that starting with some number $m^\prime+1$, all the dimensionless scattering rates $\gamma_m = r_b/(v_F\tau_m)$ (see also Eq.~\eqref{I-expand}) are the same:
\begin{equation}
    \gamma_{m>m^\prime} = \gamma=  \gamma^{(\mathrm{mr})} + \gamma_{e}^{(\mathrm{ee})}.
\end{equation}
To describe the scattering rates at $m\leqslant m^\prime$, it is convenient to introduce the deviations
\begin{equation}
    \Delta \gamma_m = \gamma - \gamma_m.
\end{equation}
In our case, $\Delta \gamma_0 =\gamma$, $\Delta \gamma_1 = \gamma_{e}^{(\mathrm{ee})}$, which follows from the charge and momentum conservation respectively, and all other even deviations are zero:
\begin{equation}
    \Delta\gamma_{2k} = 0, \qquad k\geqslant 1.
\end{equation}
However, we can also consider even relaxation rates that depend on the mode number for a limited number of even modes by substituting non-zero $\Delta\gamma_{2k}$.

Using deviations, we can recast the expansion of the angular kernel~\eqref{I-expand} as
\begin{widetext}
\begin{equation}
    \frac{r_b}{v_F} I(\theta-\theta^\prime) = \gamma\lsum_{m=0}^{+\infty} \frac{\cos{m(\theta-\theta^\prime)}}{2^{\delta_{m,0}}\pi} - \lsum_{m=0}^{m^\prime} \frac{\Delta\gamma_m}{2^{\delta_{m,0}}\pi} \cos{m(\theta-\theta^\prime)}=\\
    \gamma \delta^{(c)}(\theta-\theta^\prime) - \lsum_{m=0}^{m^\prime} \frac{\Delta\gamma_m}{2^{\delta_{m,0}}\pi} \left[\cos{m\theta}\cos{m\theta^\prime} + \sin{m\theta}\sin{m\theta^\prime}\right],
\end{equation}
\end{widetext}
where $\delta^{(c)}(\theta-\theta^\prime)$ denotes the Dirac delta-function on a circle.

Introducing dimensionless units $\rho = r/r_b$, we can then rewrite Eq.~\eqref{lboltzmann-simplified} as
\begin{multline}
    \left[\sin{\theta}\frac{\partial}{\partial \rho} + \left(\frac{\cos{\theta}}{\rho} - \rho_L^{-1}\right)\frac{\partial}{\partial\theta}\right]\eta + \gamma \eta = \\ =\sum_{m=0}^{m^\prime} \Delta\gamma_m \left[\eta_m^{(c)}\cos{m\theta}+\eta_m^{(s)}\sin{m\theta}\right],\label{lboltzmann-split}
\end{multline}
where $\rho_L = R_L/r_b$ is the dimensionless Larmor radius.

If we forget about the boundaries for a moment, the electron trajectories in the presence of the magnetic field constitute circles of radius $\rho_L$ in dimensionless units, which the electrons traverse in the clockwise direction. To simplify Eq.~\eqref{lboltzmann-split}, it is convenient to introduce new variables $(u,\alpha)$, where $u$ is the distance from the center of a circular trajectory $O^\prime$ to the center of the Corbino disk $O$, and $\alpha$ is the angular coordinate on the circular trajectory (see Fig.~\ref{trajectory-sketch}). The latter is defined in such a way that $\alpha=0$ corresponds to the point of the circular trajectory closest to the Corbino disk center $O$. For $\alpha<\pi$ (see Fig.~\ref{trajectory-sketch} $(a,b)$), $\alpha=\angle OO^\prime C$, where $C$ is the point where we want to analyze the distribution function, while for $\alpha>\pi$ (see Fig.~\ref{trajectory-sketch} $(c,d)$), $\alpha = 2\pi - \angle OO^\prime C$.
\begin{align}
    u &= \sqrt{\rho^2+\rho_L^2+2\rho\rho_L\cos{\theta}},\\
    \angle OO^\prime C &= \arccos{\frac{u^2+\rho_L^2-\rho^2}{2u\rho_L}}
    ,\label{alphadef}
\end{align}
Here, we have used the law of cosines twice.

After the coordinate change $(r,\theta)\rightarrow(u,\alpha)$, Eq.~\eqref{lboltzmann-split} becomes
\begin{equation}
    \frac{\partial\eta}{\partial \alpha} + p\eta = \rho_L\sum_{m=0}^{m^\prime} \Delta\gamma_m \left[\eta_m^{(c)}\cos{m\theta}+\eta_m^{(s)}\sin{m\theta}\right],\label{lboltzmann-split2}
\end{equation}
where we have introduced
\begin{equation}
    p = \gamma\rho_L = R_L/l.
\end{equation}

If we denote the right-hand side of Eq.~\eqref{lboltzmann-split2} as $\rho_L\chi(u,\alpha)$, we can formally write the solution as
\begin{equation}
    \eta(u,\alpha) = \BD(u)e^{-p\alpha} + \int_{\alpha_0}^{\alpha} d\alpha^\prime e^{-p(\alpha-\alpha^\prime)}\rho_L\chi(u,\alpha^\prime).
\end{equation}
It is convenient to treat the outward- and inward-moving electrons separately. From elementary geometric considerations, one can deduce that for $\alpha\in[0,\pi]$, $\theta \in[0,\pi]$ as well. In addition to that, mirror reflecting the point $C$ with respect to the line $OO^\prime$ transforms $(\theta,\alpha)\rightarrow(2\pi-\theta,2\pi-\alpha)$ (see Fig.~\ref{trajectory-sketch}). As a result, we can identify the angles $\alpha\in(0,\pi)$ ($\alpha\in(\pi,2\pi)$) with the outward-(inward-)moving electrons. We can also deduce that the specular reflection at the boundary changes the angle as $\alpha\rightarrow(2\pi-\alpha)$, but does not affect $u$.

In order to find $\BD(u)$, we substitute the solution into the boundary conditions~\eqref{inner-boundary} and~\eqref{outer-boundary} and solve the resulting system of equations. This way, we obtain
\begin{widetext}
\begin{multline}
    d\times\eta(u,\alpha) = (1-r_{\theta_a})e\Phi_a \times e^{-p(\alpha-\alpha_a)} +(1-r_{\theta_b})e\Phi_b \times r_{\theta_a}e^{-p(\alpha_b-\alpha_a)}e^{-p(\alpha-\alpha_a)} + \lint_{\alpha_{a}}^\alpha e^{-p(\alpha-\alpha^\prime)}\chi(u, \alpha^\prime)\rho_Ld\alpha^\prime
    +\\+ r_{\theta_a}r_{\theta_b} e^{-p(\alpha_b-\alpha_a)}\int_{\alpha}^{\alpha_b} e^{-p(\alpha-\alpha_a+\alpha_b-\alpha^\prime)}\chi(u,\alpha^\prime)\rho_Ld\alpha^\prime
    + r_{\theta_a}\lint_{2\pi-\alpha_b}^{2\pi-\alpha_a} e^{-p\left[(\alpha-\alpha_a) + (2\pi-\alpha_a-\alpha^\prime)\right]}\chi(u,\alpha^\prime)\rho_Ld\alpha^\prime,\label{outward-distribution}
\end{multline}
\begin{multline}
d\times\eta(u,2\pi-\alpha) = (1-r_{\theta_b}) e\Phi_b \times e^{-p(\alpha_b-\alpha)} + (1-r_{\theta_a})e\Phi_a\times r_{\theta_b}e^{-p(\alpha_b-\alpha_a)}e^{p(\alpha_b-\alpha)} + \lint_{2\pi-\alpha_b}^{2\pi-\alpha} e^{p(2\pi-\alpha-\alpha^\prime)}\chi(u,\alpha^\prime)\rho_Ld\alpha^\prime +\\+
r_{\theta_a}r_{\theta_b} e^{-p(\alpha_b-\alpha_a)}\lint_{2\pi-\alpha}^{2\pi-\alpha_a} e^{-p\left[(\alpha_b-\alpha)+(2\pi-\alpha_a-\alpha^\prime)\right]}\chi(u,\alpha^\prime)\rho_Ld\alpha^\prime+
r_{\theta_b}\lint_{\alpha_a}^{\alpha_b}e^{-p(\alpha_b-\alpha+\alpha_b-\alpha^\prime)}\chi(u,\alpha^\prime)\rho_Ld\alpha^\prime,\label{inward-distribution}
\end{multline}
\end{widetext}
where
\begin{equation}
    d = 1 - r_{\theta_a}r_{\theta_b}e^{-2p(\alpha_b-\alpha_a)}.
\end{equation}

The angles $\alpha_a$ and $\alpha_b$ correspond to the points where the circular trajectory crosses the boundary:
\begin{align}
    \alpha_a &= \arccos{\left[\min{\left\{\frac{u^2+\rho_L^2-\rho_a^2}{2u\rho_L}, 1\right\}}\right]},\\
    \alpha_b & = \arccos{\left[\max{\left\{\frac{u^2+\rho_L^2-1}{2u\rho_L}, -1\right\}}\right]}.
\end{align}
Here, $\rho_a = r_a/r_b$, and we took into account that $\rho_b = r_b/r_b=1$.
The angles $\theta_a$ and $\theta_b$ correspond to the same crossing points as well:
\begin{align}
    \theta_a &= \arccos{\left[\min{\left\{\frac{u^2-\rho_L^2-\rho_a^2}{2\rho_a\rho_L}, 1\right\}}\right]},\\
    \theta_b & = \arccos{\left[\max{\left\{\frac{u^2-\rho_L^2-1}{2\rho_L}, -1\right\}}\right]}.
\end{align}
If the trajectory does not cross the inner boundary, then $\alpha_a=\theta_a=0$. Correspondingly, if it does not touch the outer one, then $\alpha_b=\theta_b = \pi$.

Despite their formidable look, the form of Eqs.~\eqref{outward-distribution} and~\eqref{inward-distribution} can be deduced purely from physical arguments. Equation~\eqref{lboltzmann-split2} can be interpreted in the following manner: The right-hand side introduces deformation of the distribution function, while the left-hand side describes how this deformation is carried by the electrons along the trajectory and exponentially decays as $e^{-p\Delta\alpha}$, where $\Delta\alpha$ is the angular distance covered along the trajectory. When the electrons scatter at the boundary, the deformation is propagated further only if they reflect specularly, which introduces another probability factor $r_\theta$. Overall, the distribution function at the specific point can be represented as the sum of weighted contributions of all the parts of the trajectory.

Let us analyze the case of the outward-moving electrons (Eq.~\eqref{outward-distribution}) in more detail. The first two terms describe the boundary contributions: deformation from the inner boundary travels directly to the point with angular coordinate $\alpha$, while the deformation from the outer boundary has to travel to the inner boundary and get reflected once, hence the additional $r_{\theta_a}e^{-p(\alpha_b-\alpha_a)}$ probability factor. The third term describes the contribution of the outward-moving electrons arriving from smaller angles $\alpha^\prime<\alpha$. At the same time, the outward-moving electrons with larger angles $\alpha^\prime>\alpha$ have to reflect at both boundaries to arrive in the point of the trajectory corresponding to the angle $\alpha$, which is reflected in the fourth term. The fifth term describes the contribution of the inward-moving electrons, which have to reflect once at the inner boundary. Finally, we also need to take into account that the electrons can loop several times around the trajectory~\footnote{Due to the axial symmetry, we can identify as equivalent all the points of the Corbino disk at the same distance from the center. In this sense, all the trajectories are closed and periodic even despite the reflection at the boundaries, because the latter does not change parameter $u$.}, which is covered by the factor
\begin{multline}
    1/d = 1 + r_{\theta_a}r_{\theta_b}e^{-2p(\alpha_b-\alpha_a)} + \left(r_{\theta_a}r_{\theta_b}e^{-2p(\alpha_b-\alpha_a)}\right)^2 +\\+ \left(r_{\theta_a}r_{\theta_b}e^{-2p(\alpha_b-\alpha_a)}\right)^3 +\dotsc,
\end{multline}
where $r_{\theta_a}r_{\theta_b}e^{-2p(\alpha_b-\alpha_a)}$ is precisely the probability that the deformation propagates one period along the trajectory. Note that we have multiplied both sides by $d$, which is why $d$ appears in the left-hand side of Eq.~\eqref{outward-distribution}. The case of the inward-moving electrons (Eq.~\eqref{inward-distribution}) can be analyzed in an analogous manner.

If the trajectory does not cross the inner (outer) boundary, then the corresponding probability of reflection that enters Eqs.~\eqref{outward-distribution} and~\eqref{inward-distribution} must be $r_{\theta_a}=1$ ($r_{\theta_b}=1$). So, for the equations to have the correct physical interpretation, it is crucial to put $r_0=r_\pi =1$, even if we consider fully diffusive boundary conditions ($r_\theta\equiv0$).

To find the closed system of integral equations, we project Eqs.~\eqref{outward-distribution} and~\eqref{inward-distribution} onto the first $m^\prime+1$ angular harmonics using
\begin{multline}
    \eta_k^{(c)}(\rho,\theta) = \frac{1}{2^{\delta_{k,0}}\pi}\lint_{0}^{2\pi} \eta(\rho,\theta)\cos{k\theta} d\theta =\\= \frac{1}{2^{\delta_{k,0}}\pi}\lint_0^{\pi}\left[\eta(\rho,\theta)+\eta(r,2\pi-\theta)\right]\cos{k\theta}d\theta.
\end{multline}
\begin{multline}
    \eta_k^{(s)}(r,\theta) = \frac{1}{\pi}\lint_{0}^{2\pi} \eta(r,\theta)\sin{k\theta} d\theta =\\= \frac{1}{\pi}\lint_0^{\pi}\left[\eta(r,\theta)-\eta(r,2\pi-\theta)\right]\sin{k\theta}d\theta.
\end{multline}
The end result is
\begin{widetext}
\begin{equation}
    \eta_{k}^{(r)}(\rho) = e\Phi_a\lint_{0}^\pi d\theta\BA_k^{(r)}(\rho,\theta) + e\Phi_b \lint_{0}^{\pi}d\theta\BB_k^{(r)}(\rho,\theta) + \sum_{k^\prime=0}^{m^\prime} \Delta\gamma_{k^\prime} \sum_{r^\prime=c,s}\lint_{0}^{\pi}d\theta\lint_{\alpha_{a}(\rho,\theta)}^{\alpha_b(\rho,\theta)} \rho_L d\alpha^\prime G_{k,k^\prime}^{r,r^\prime}(\rho,\theta,\alpha^\prime) \eta_{k^\prime}^{r^\prime}(\rho^\prime(\rho,\theta,\alpha^\prime)),\label{integral-equation}
\end{equation}
\end{widetext}
where the index $r=c,s$ labels the cosine and sine harmonics. The boundary kernels are
\begin{multline}
    \BA_k^{(r)}(\rho,\theta)=\\\frac{1-r_{\theta_a}}{2^{\delta_{k,0}}\pi d}\left[e^{-p(\alpha-\alpha_a)}+\sigma_r r_{\theta_b}e^{-p(2\alpha_b-\alpha_a-\alpha)}\right] g_k^{(r)}(\theta),
\end{multline}
\begin{multline}
    \BB_k^{(r)}(\rho,\theta)=\\\sigma_r\frac{1-r_{\theta_b}}{2^{\delta_{k,0}}\pi d}\left[e^{-p(\alpha_b-\alpha)}+\sigma_r r_{\theta_a}e^{-p(\alpha+\alpha_b-2\alpha_a)}\right] g_k^{(r)}(\theta),
\end{multline}
where
\begin{equation}
\begin{array}{ccc}
    g_k^{(c)} = \cos{k\theta}, & \quad& \sigma_c = +1,\\
    g_k^{(s)} = \sin{k\theta}, & \quad& \sigma_s = -1.
\end{array}
\end{equation}
The bulk kernels are
\begin{widetext}
\begin{multline}
    G_{k,k^\prime}^{r,r^\prime}(\rho,\theta,\alpha^\prime) = 
    \frac{g_k^{(r)}(\theta)g_{k^\prime}^{(r^\prime)}(\theta^\prime)}{2^{\delta_{k,0}}\pi d}\left\{f_{\sigma_r\sigma_r^\prime}(\alpha-\alpha^\prime)\left[e^{-p|\alpha-\alpha^\prime|}+ \sigma_r\sigma_{r^\prime}r_{\theta_a}r_{\theta_b}e^{-2p(\alpha_b-\alpha_a) + p|\alpha-\alpha^\prime|}\right] +\right.\\+\left. \sigma_r r_{\theta_b}e^{-p(2\alpha_b - \alpha-\alpha^\prime)}+\sigma_{r^\prime}r_{\theta_a}e^{-p(\alpha+\alpha^\prime-2\alpha_a)}\right\},
\end{multline}
\end{widetext}
where $f_{+1}(\alpha-\alpha^\prime)\equiv1$ and $f_{-1}(\alpha-\alpha^\prime) = \sign{(\alpha-\alpha^\prime)}$.

A comment is in order about the functional dependence of all these parameters on each other. The coordinates $(\rho,\theta)$ determine the specific trajectory and all its corresponding parameters: $u, \alpha_{a,b}, \theta_{a,b}$. They also determine the specific point on the trajectory where the distribution is analyzed and hence the angle $\alpha$. The coordinates $(\rho^\prime, \theta^\prime)$ correspond to other points of the trajectory that are determined through the integration variable $\alpha^\prime$:
\begin{align}
    \rho^\prime &= \sqrt{u^2+\rho_L^2-2u\rho_L\cos{\alpha^\prime}},\\
    \theta^\prime & = \arccos{\frac{u^2-\rho_L^2-{\rho^\prime}^2}{2\rho_L\rho^\prime}}.
\end{align}

The form of the system of integral equations in the absence of a magnetic field can be obtained directly by taking the limit $\rho_L\rightarrow+\infty$ in all the equations.
We can achieve this by replacing
\begin{equation}
    p\alpha = \gamma\rho_L\alpha\rightarrow \gamma w,
\end{equation}
where $w = \rho\sin{\theta}$ is the distance along the trajectory in the dimensionless units (see the geometrical sketch in Fig.~\ref{zero-field-sketch}).
In the same spirit we need to replace $p\alpha_{a,b}\rightarrow w_{a,b}/l$,
where~\footnote{In the limit of zero magnetic field, the trajectories in the absence of boundaries become straight lines. They may not cross the inner lead, however, they always cross the outer lead.}
\begin{align}
w_a & = \sqrt{\max{\{\rho_a^2-\rho^2\cos^2{\theta}, 0\}}},\\
w_b & = \sqrt{\rho_b^2 - \rho^2\cos^2{\theta}}.
\end{align}
The equations for the angles $\theta_{a,b}$ take the form
\begin{align}
\theta_a &= \arccos{\min{\left\{\frac{\rho}{\rho_a}\cos{\theta}, 1\right\}}},\\
\theta_b & = \arccos{\frac{\rho}{\rho_b}\cos{\theta}}.
\end{align}
Finally, the integration variable in the Eq.~\eqref{integral-equation} needs to be replaced $\rho_L d\alpha^\prime\rightarrow dw^\prime$ together with the integration limits $\alpha_{a,b}\rightarrow w_{a,b}$. The coordinates $(\rho^\prime,\theta^\prime)$ on the trajectory can then be expressed as
\begin{align}
    \rho^\prime & = \sqrt{{w^\prime}^2 + \rho^2\cos^2{\theta}},\\
    \theta^\prime & = \arccos{\frac{\rho}{\rho^\prime}\cos{\theta}}.
\end{align}

Another simplifying property in the absence of a magnetic field is that all the $\cos{m\theta}$ harmonics are strictly zero, which follows from the fact that mirror reflection with respect to a diagonal of the Corbino disk is then a symmetry of the problem.

\subsection{Details of the numeric implementation}

All the expressions for the angles defined via $\arccos$  used throughout the section were numerically unstable.
We have employed them because it is easier to deduce their geometrical meaning.
In the practical realization, we transformed all the expressions using the fact that
\begin{equation}
    \alpha = 2\arctan{\sqrt{\frac{1-\cos{\alpha}}{1+\cos{\alpha}}}}, \qquad 0\leqslant\alpha\leqslant \pi.
\end{equation}
We then used the stable formula described in Ref.~\cite{Kahan_2014}.

Due to current conservation, $\rho\eta_1^{(s)}(\rho)\equiv C$, which we could have substituted explicitly into the integral equation as it has been done in Ref.~\cite{Raichev_2022}. We have decided against it because it would complicate the code implementation. In addition to that, if we do not enforce current conservation, we can use it as an internal check of the method. Specifically, we can use the quantity
\begin{equation}
\frac{\mathrm{Standard\  Error}\left[\rho\eta_1^{(s)}(\rho)\right]}{\mathrm{Mean} \left[\rho\eta_1^{(s)}(\rho)\right]}
\end{equation}
as an estimate of the relative error.

To obtain the proper system of integral equations, we need to make the change of integration variable $\alpha^\prime\rightarrow \rho^\prime$ ($w^\prime\rightarrow \rho^\prime$) in the presence (absence) of a magnetic field. However, the resulting bulk kernels have a logarithmic singularity at $\rho-\rho^\prime=0$. So, for practical purposes it was actually more convenient to keep the system in the form~\eqref{integral-equation}. We discretized the integral equations by approximating $\eta_k^{(r)}(\rho)$ using a piecewise continuous polynomial basis and then evaluating all the integrals numerically (see Section 3.4.4 of Ref.~\cite{Atkinson_1997} and references therein). The idea is to split the interval $[\rho_a,\rho_b]$ into smaller ones, and then, for every smaller interval, use the basis of Lagrange interpolating polynomials collocated at Gauss-Legendre points to approximate the components of the distribution function.

To determine the interval splitting, we took into account the fact that the solutions of the integral equations can develop jumps in the first derivative at specific values of $\rho$: $|2\rho_L-1|$, $|2\rho_L \pm \rho_a|$~\footnote{These values correspond to the extremal points of the circular trajectories that touch either of the leads.}. We will refer to these values as breaking points.
Let $\rho_i$, $i=1,2,\dotsc,s$, be a set of ordered dimensionless radii, where $\rho_1 = \rho_a$, $\rho_s=\rho_b=1$ and $\rho_i$ for $1 < i < s$ correspond to those breaking points that belong to the interval $[\rho_a, 1]$~\footnote{We assume all the radii to be unique. If some of the breaking points coincide together or with $\rho_a$ or $\rho_b=1$, we count them only once.}.

The intervals we used were then determined in the following way. We have split every range $[\rho_i, \rho_{i+1}]$ using the Chebyshev nodes
\begin{multline}
t^{(i)}_l = \frac{\rho_{i}+\rho_{i+1}}{2} - \frac{\rho_{i+1}-\rho_i}{2}\cos{\frac{\pi(2l+1)}{2n_i+2}},\\ l=0,1,\dotsc,n_i
\end{multline}
Here, $n_i$ was determined as
\begin{equation}
    n_i = \left\lceil\frac{N}{m}\frac{\rho_{i+1}-\rho_i}{1-\rho_a}\right\rceil,
\end{equation}
where $m$ is the number of Gauss-Legendre nodes within each of the intervals. In the computations we have used $m=3$. $N$ is the parameter that sets the lower bound for the overall number of the components of the discretized distribution function. In the computations we kept $N=400$. The intervals themselves were $[\rho_i, t^{(i)}_{0}],[t^{(i)}_{0}, t^{(i)}_{1}], [t^{(i)}_{1}, t^{(i)}_{2}],\dotsc, [t^{(i)}_l,\rho_{i+1}]$ for all $i=1,2,\dotsc, s$.

The number of kernel components grows quadratically with the number of long-lived harmonics we keep. To keep the computation time manageable, we had to restrict the number of long-lived harmonics to $5$ in the presence of a magnetic field and $8$ in its absence.

\section{Numerical method based on direct discretization of the linearized Boltzmann equation\label{details-discretization}}

Using the dimensionless units $\rho=r/r_b$ introduced in Appendix~\ref{details-integral-eq}, we write the linearized Boltzmann equation in the form
\begin{equation}
    \left[\sin{\theta}\frac{\partial}{\partial \rho} + \left(\frac{\cos{\theta}}{\rho} - \rho_L^{-1}\right)\frac{\partial}{\partial\theta} + \hat I_c\right]\eta = 0,
\end{equation}
where
\begin{multline}
\hat I_c \sum_{m}[\eta_m^{(c)}\cos{m\theta} + \eta_m^{(s)}\sin{m\theta}] \\= \sum_m \gamma_m [\eta_m^{(c)}\cos{m\theta} + \eta_m^{(s)}\sin{m\theta}].
\end{multline}

To discretize the equation, we collocated the distribution function at dimensionless radii
\begin{equation}
    \rho_i = \rho_a + \frac{1-\rho_a}{N_\rho}(i-1/2),\qquad i=1,2,\dotsc, N_\rho
\end{equation}
and angles
\begin{equation}
    \theta_j = \frac{\pi(j - 1/2)}{N_\theta},\qquad j= 1,2,\dotsc ,2N_\theta.
\end{equation}
The partial derivatives were then approximated as central differences in the bulk:
\begin{align}
    \partial\eta/\partial\rho&\rightarrow \Delta_\rho \eta^i_j = \frac{\eta^{i+1}_j-\eta^{i-1}_j}{2\Delta\rho},\\
    \partial\eta/\partial\theta&\rightarrow\Delta_\theta \eta^i_j = \frac{\eta^{i}_{j+1}-\eta^{i}_{j-1}}{2\Delta\theta}.
\end{align}
Here, $\Delta\rho = (1-\rho_a)/N_\rho$, $\Delta\theta = \pi/N_\theta$.

The partial derivatives with respect to $\rho$ at the boundaries required separate treatment.
Let us discuss the case of the inner boundary in detail. For the inflow electrons we have used the one-sided finite difference scheme of the second order:
\begin{equation}
\Delta_\rho \eta^{1}_j = \frac{3\eta^1_j - 4\eta^2_j+\eta^3_j}{2\Delta\rho},\qquad j=N_\theta+1,N_\theta+2,\dotsc 2N_\theta.
\end{equation}
For the outflow electrons, we first extrapolated the distribution function of the inflow electrons to the boundary:
\begin{multline}
    \eta^{1/2}_{2N_\theta -j} = \frac{15}{8}\eta^1_{2N_\theta -j}  - \frac{5}{4}\eta^2_{2N_\theta -j}  + \frac{3}{8}\eta^3_{2N_\theta -j},\\j=1,2,\dotsc, N_\theta,
\end{multline}
and applied the boundary condition~\eqref{inner-boundary}
\begin{equation}
    \eta^{1/2}_j = (1-r_{\theta_j})e\Phi_a + r_{\theta_j} \eta^{1/2}_{2N_\theta-j}.
\end{equation}
Finally, we approximated the partial derivative using a second-order scheme
\begin{equation}
    \Delta_\rho \eta^{1}_j = \frac{\eta^2_j + 3 \eta^1_j - 4\eta^{1/2}_j}{3\Delta\rho}.
\end{equation}
The outer boundary was treated analogously.

To efficiently compute the action of the linearized collision operator $\hat I_c$, we did discrete Fourier transform, multiplied with the dimensionless collision rates $\gamma_m$ and then performed the inverse discrete Fourier transform.

As we see, the boundary potentials enter only through the discretization of $\partial\eta/\partial\rho$ at the boundaries. If we move these terms to the right-hand side, we obtain a system of linear equations
\begin{equation}
\hat L \bar \eta = \bar b,
\end{equation}
which can be solved using standard iterative methods.
Here, $\bar \eta = \eta^i_j$ denotes the discretized distribution function. Notice that $\bar b$ is zero everywhere within the bulk.

We have programmed the discretization of the linearized Boltzmann equation using the Julia programming language~\cite{Bezanson_2017}. The code was designed to run on NVIDIA GPUs using the CUDA.jl library~\cite{Besard_2019}. We then solved the resulting linear system of equations using the restarted version of the GMRES method~\cite{Saad_1986}. Specifically, we employed the implementation in Krylov.jl library~\cite{Montoison_2023}.

In the computations, we have kept $N_\rho=4000$ and $N_\theta=225$ for the temperature run in the absence of a magnetic field and $N_\rho=1000$ and $N_\theta=1600$ in the presence of a magnetic field. The restarted GMRES solver was run until the absolute tolerance of $10^{-6}||\bar b||$ was reached. To speed up the process, the solution for the previous set of parameters was used as the seed of the iterative procedure for the new slightly altered set of parameters.

It is interesting to note that in the presence of a magnetic field, solutions tend to develop oscillations that disappear only deep into the hydrodynamic regime. It seems to be the result of the interplay of the boundary conditions with the magnetic field: the distribution function experiences jumps at $\theta=0,\pi$ at the boundaries; when the distribution is propagated inside the bulk, the magnetic field rotates these jumps to finite angles.
The only practical way to kill off these oscillations that we were able to find was to increase the angular resolution of the mesh, which explains the value of $N_\theta$ used for parameter runs in the presence of the magnetic field.


\bibliography{bibliography.bib, supplement.bib}

\clearpage

\newcommand{\beginsupplement}{%
        \setcounter{table}{0}
        \renewcommand{\thetable}{S\arabic{table}}%
        \setcounter{figure}{0}
        \renewcommand{\thefigure}{S\arabic{figure}}%
        \setcounter{equation}{0}
        \renewcommand{\theequation}{S\arabic{equation}}
        \setcounter{section}{0}
        \renewcommand{\thesection}{S\Roman{section}}
     }

\onecolumngrid
\begin{center}
    \bf\large
Supplementary Material\\ for the article\\ ``Tomographic flow regime vs even-odd effect for the magnetotransport in the Corbino geometry''
\end{center}


\beginsupplement

All equation numbers, figure numbers and reference numbers without the prefix ``S'' refer to the respective numbers in the main article.

\section{Estimates of the experimental parameters in Ref.~\cite{Zeng_2024}}

For the estimates, we focus on the temperature range $50-200\ \mathrm{K}$, in which the results attributed to the tomographic flow regime were reported in Ref.~\cite{Zeng_2024}.

The scattering rate for the momentum-relaxing process can be derived from the provided data on the resistance sensitivity in high magnetic fields (Fig. 1e of Ref.~\cite{Zeng_2024}),
\begin{equation}
    \left.\frac{\partial R}{\partial(B^2)}\right|_{\mathrm{large\ } B} = \alpha^{Ohm} = \frac{\ln{r_b/r_a}}{2\pi \rho_o(ne)^2},
\end{equation}
where, in the case of monolayer graphene, the Ohmic resistivity $\rho_o$ is given by
\begin{equation}
    \rho_o = \frac{m^*}{ne^2 \tau^{(\mathrm{mr})}} = \frac{\hbar k_F/v_F}{ne^2 \tau^{(\mathrm{mr})}} = \frac{\hbar\sqrt{\pi n}/v_F}{ne^2 \tau^{(\mathrm{mr})}}
\end{equation}
This way, we can express the corresponding scattering length $l^{(\mathrm{mr})}=v_F\tau^{(\mathrm{mr})}$ as
\begin{equation}
    l^{(\mathrm{mr})} = \frac{h\sqrt{\pi n^3}}{\ln{r_b/r_a}} \alpha^{Ohm}.
\end{equation}
For the temperatures $50-200\ \mathrm{K}$, the reported $\alpha^{Ohm}$ changed in the range $60-30\ \mathrm{ k\Omega/T^2}$ in the sample with $n=0.2\times10^{12}\ \mathrm{cm^{-2}}$, which corresponds to the scattering lengths $4.6-2.3\ \mathrm{\mu m}$. For the sample with the smallest $r_b=1.75\ \mathrm{\mu m}$, this corresponds to the dimensionless scattering rates $\gamma^{(\mathrm{mr})} = r_b/l^{(\mathrm{mr})}\sim 0.38-0.77$, while for the sample with the largest $r_b=4.5\ \mathrm{\mu m}$ we get $\gamma^{(\mathrm{mr})} \sim 0.99-1.98$.

As we have mentioned in the main text, the electron-electron scattering rates depend on the dimensionless interaction strength $r_s = 1/(\sqrt{2}k_F l_{TF})$, where the Thomas-Fermi screening length is given by
\begin{equation}
    l_{TF} = \frac{\varepsilon}{2\pi e^2 \nu_F} = \frac{\varepsilon \hbar^2}{g_sg_v m^* e^2}.
\end{equation}
Here, $\varepsilon$ is the effective dielectric constant induced by the substrate. In the experiment, the sheet of mono- or bilayer graphene was sandwiched between hexagonal Boron Nitride (hBN), so we take $\varepsilon=\sqrt{\varepsilon_\parallel\varepsilon_\perp}$~\cite{Kim_2020}, where $\varepsilon_{\parallel,\perp}$ are the in- and out-of-plane dielectric constants for hBN. For this estimate, we take the bulk values~\cite{Laturia_2018} $\varepsilon_\parallel= 6.93$ and $\varepsilon_\perp = 3.76$.

Since for monolayer graphene $m^* = \hbar k_F/v_F$, the dimensionless scattering strength does not depend on the density:
\begin{equation}
   r_s = \frac{g_sg_v}{\sqrt{2}\varepsilon} \alpha^*=1.21,
\end{equation}
where $\alpha^* = e^2/\hbar v_F \approx 2.2$ is the fine-structure constant for graphene.

For bilayer graphene, $m^*/m_e = 0.03$ and it is convenient to rewrite the dimensionless scattering strength as
\begin{equation}
    r_s = \frac{m^*}{m_e} \frac{g_s g_v}{\varepsilon} \frac{1}{a_B\sqrt{2\pi n}},
\end{equation}
where $a_B = \hbar^2/(m_e e^2)$ is the Bohr radius.
The experiments with bilayer graphene were performed for different densities of the charge carriers. We list the corresponding densities and the computed values of $r_s$ in Table~\ref{table:rs-bilayer}
\begin{table}[h]
\caption{Estimated values of $r_s$ for the experiments presented in Fig. 3(d) of Ref.\cite{Zeng_2024}.}
\label{table:rs-bilayer}
\begin{tabular}{|c|c|c|c|}\hline
$n,\ 10^{12}\ \mathrm{cm^{-2}}$ & $0.61$ & $1.01$ & $1.81$ \\\hline
$r_s$ & $2.27$ & $1.76$ & $1.32$ \\\hline
\end{tabular}
\end{table}

In order to compute the electron-electron scattering rates, it is convenient to introduce the dimensionless screened Coulomb interaction,
\begin{equation}
    \bar V(q) = \frac{m^*}{2\pi\hbar^2} \frac{2\pi e^2}{\varepsilon(q + l_{TF}^{-1})} = \frac{1}{g_sg_v}\frac{r_s}{r_s+\sqrt{2}\frac{q}{2k_F}}
\end{equation}
Thus, we find the dimensionless scattering rate for the odd harmonics as~\cite{Nilsson_2025}
\begin{equation}
    \gamma_o^{(\mathrm{ee})} = \frac{2\pi^3 E_F r_b}{15\hbar v_F} W^2 \left(\frac{T}{E_F}\right)^4,
\end{equation}
where $W^2$ is the dimensionless squared antisymmetrized vertex, resummed over spin and valley indices,
\begin{equation}
    W^2 = (2g_v-1) [\bar V^2(0) + \bar V^2(2k_F)] + [\bar V(0)-\bar V(2k_F)]^2 = 2g_v[\bar V^2(0) + \bar V^2(2k_F)] - 2\bar V(0)\bar V(2k_F).
\end{equation}
Here, we have explicitly taken into account $g_s = 2$, since the expression is meaningless in the spinless case. If $g_v=1$, $W^2$ corresponds to the normalized $|V^*|^2$ used in Ref.~\cite{Ledwith_2019a}.
Substituting the dimensionless form of the interaction explicitly, we obtain
\begin{equation}
    W^2 = \frac{1}{2g_v}\left[1 + \frac{r_s}{g_v}\frac{(g_v-1)r_s-\sqrt{2}}{(r_s+\sqrt{2})^2}\right]
\end{equation}

For the even harmonics, we need to take into account that the valley degeneracy changes the weight of the direct contribution. As a result, we obtain~\cite{Nilsson_2025}
\begin{equation}
    \gamma_{e}^{(\mathrm{ee})} = \frac{4\pi E_F r_b}{3\hbar v_F} \left(\frac{T}{E_F}\right)^2 [g_v I^{\mathrm{dir}} + I^{\mathrm{ex}}],
\end{equation}
where
\begin{align}
    I^{\mathrm{dir}} = r_s^2\left[\ln\left(1 + \frac{\sqrt{2}}{r_s}\right) - \frac{\sqrt{2}}{\sqrt{2} +r_s}\right].
\end{align}
As the authors of Ref.~\cite{Nilsson_2025} point out, $I^{\mathrm{ex}} \approx -I^\mathrm{dir}/2$ for $r_s\gtrsim0.5$ with good accuracy, which is well satisfied in our case.

Finally, we can estimate the phenomenological parameter $a$ as
\begin{equation}
    a = \frac{\left(\gamma_{e}^{(\mathrm{ee})}\right)^2}{\gamma_o^{(\mathrm{ee})}} = \frac{40}{3\pi}\frac{E_Fr_b}{\hbar v_F} \frac{\left[(2g_v-1)I^\mathrm{dir}/2\right]^2}{W^2}.~\label{eq:a-estimate}
\end{equation}
Here, $E_Fr_b/(\hbar v_F) = k_Fr_b/\delta$, where $\delta=1$ for the linear spectrum and $\delta=2$ for the parabolic spectrum. We provide the results of the computation using Eq.~\eqref{eq:a-estimate} in Tables~\ref{table:a-mlg} and~\ref{table:a-blg} for the cases of mono- and bilayer graphene respectively.

\begin{table}[H]
\centering
\caption{Estimates of the parameter $a$ in the MLG for the parameters listed in Fig. 3(c) of Ref.~\cite{Zeng_2024}}
\label{table:a-mlg}
\begin{tabular}{|c|c|c|}\hline
 & $r_b=4.5\ \mathrm{\mu m}$ & $r_b=1.75\ \mathrm{\mu m}$ \\ \hline
 $n = 0.34\times10^{12}\ \mathrm{cm^{-2}}$ & $2152$ & $837$ \\\hline
 $n = 0.95\times10^{12}\ \mathrm{cm^{-2}}$ & $3596$ & $1399$ \\\hline
 $n = 1.85\times10^{12}\ \mathrm{cm^{-2}}$ & $5018$& $1952$ \\\hline
\end{tabular}
\end{table}

\begin{table}[H]
    \centering
\caption{Estimates of the parameter $a$ in the BLG for the parameters listed in Fig. 3(d) of Ref.~\cite{Zeng_2024}}
\label{table:a-blg}
\begin{tabular}{|c|c|c|}\hline
 & $r_b=4.5\ \mathrm{\mu m}$ & $r_b=1.75\ \mathrm{\mu m}$ \\ \hline
 $n = 0.61\times10^{12}\ \mathrm{cm^{-2}}$ & $2972$ & $1156$ \\\hline
 $n = 1.01\times10^{12}\ \mathrm{cm^{-2}}$ & $2972$ & $1156$ \\\hline
 $n = 1.81\times10^{12}\ \mathrm{cm^{-2}}$ & $2785$& $1083$ \\\hline
\end{tabular}
\end{table}

In addition to that, it is also interesting to estimate the even dimensionless scattering rate for a typical temperature of $T=100\ \mathrm{K}$. We present the results of the computation in Tables~\ref{table:ge-mlg} and~\ref{table:ge-blg}
\begin{table}[H]
    \centering
\caption{Estimates of the even dimensionless scattering rate at $T=100\ \mathrm{K}$ in the MLG for the parameters listed in Fig. 3(c) of Ref.~\cite{Zeng_2024}}
\label{table:ge-mlg}
\begin{tabular}{|c|c|c|}\hline
 & $r_b=4.5\ \mathrm{\mu m}$ & $r_b=1.75\ \mathrm{\mu m}$ \\ \hline
 $n = 0.34\times10^{12}\ \mathrm{cm^{-2}}$ & $16.2$ & $6.3$ \\\hline
 $n = 0.95\times10^{12}\ \mathrm{cm^{-2}}$ & $9.6$ & $3.8$ \\\hline
 $n = 1.85\times10^{12}\ \mathrm{cm^{-2}}$ & $6.9$& $2.7$ \\\hline
\end{tabular}
\end{table}

\begin{table}[H]
    \centering
\caption{Estimates of the even dimensionless scattering rate at $T=100\ \mathrm{K}$ in the BLG for the parameters listed in Fig. 3(d) of Ref.~\cite{Zeng_2024}}
\label{table:ge-blg}
\begin{tabular}{|c|c|c|}\hline
 & $r_b=4.5\ \mathrm{\mu m}$ & $r_b=1.75\ \mathrm{\mu m}$ \\ \hline
 $n = 0.61\times10^{12}\ \mathrm{cm^{-2}}$ & $126.9$ & $49.4$ \\\hline
 $n = 1.01\times10^{12}\ \mathrm{cm^{-2}}$ & $51.5$ & $20.0$ \\\hline
 $n = 1.81\times10^{12}\ \mathrm{cm^{-2}}$ & $17.6$& $6.9$ \\\hline
\end{tabular}
\end{table}

Finally, we also compute the values of the magnetic field $B_0$, at which the Larmor radius equals $r_b$, for ease of comparison with the experiment. Since the densities for the BLG and MLG were similar, we provide only the results for MLG in Table~\ref{table:b0-mlg}

\begin{table}[H]
    \centering
\caption{Values of $B_0$ in the MLG for the parameters listed in Fig. 3(c) of Ref.~\cite{Zeng_2024}}
\label{table:b0-mlg}
\begin{tabular}{|c|c|c|}\hline
 & $r_b=4.5\ \mathrm{\mu m}$ & $r_b=1.75\ \mathrm{\mu m}$ \\ \hline
 $n = 0.34\times10^{12}\ \mathrm{cm^{-2}}$ & $15.1\ \mathrm{mT}$ & $38.9\ \mathrm{mT}$ \\\hline
 $n = 0.95\times10^{12}\ \mathrm{cm^{-2}}$ & $25.3\ \mathrm{mT}$ & $65.0\ \mathrm{mT}$ \\\hline
 $n = 1.85\times10^{12}\ \mathrm{cm^{-2}}$ & $35.3\ \mathrm{mT}$& $90.7\ \mathrm{mT}$ \\\hline
\end{tabular}
\end{table}

\nopagebreak
\section{Additional plots for alternative values of $\gamma^{(\mathrm{mr})}$}

\begin{figure}[!p]
\centering
\includegraphics[width=\columnwidth]{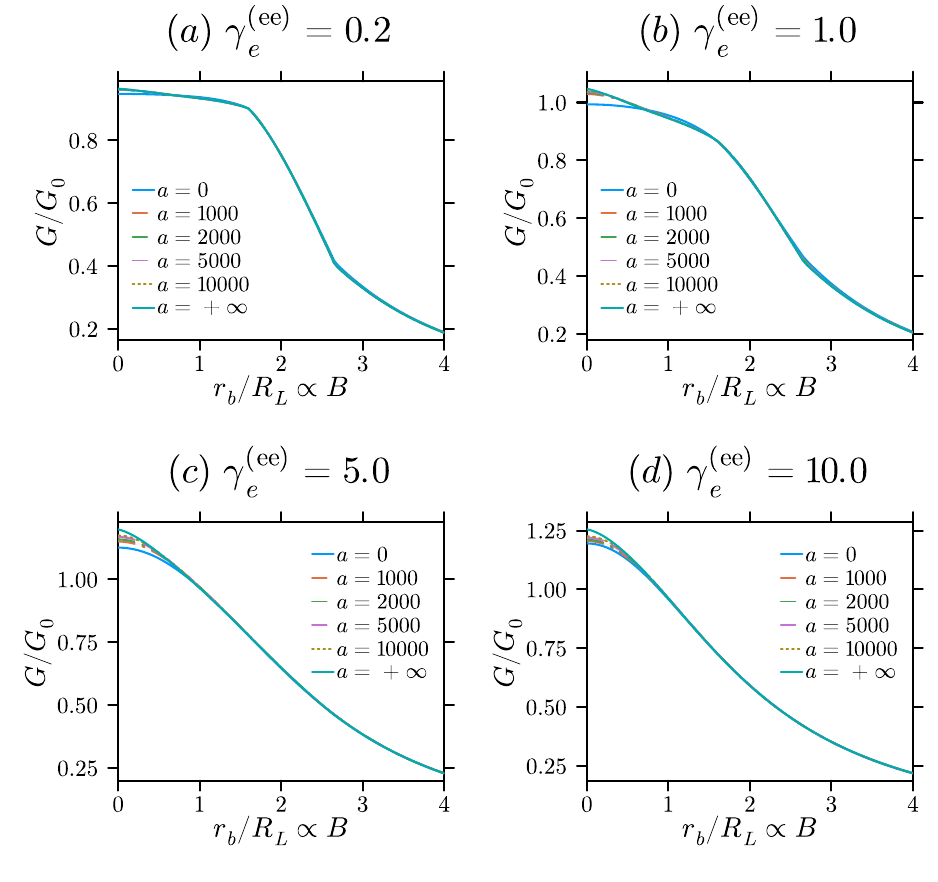}
\caption{Case $\gamma^{(\mathrm{mr})}=0.5$: Normalized total conductance $G/G_0$ of the Corbino disk as a function of the inverse Larmor radius $r_b/R_L\propto B$ for different values of the phenomenological parameter $a$ and the even electron-electron scattering rate $\gamma_{e}^{(\mathrm{ee})}$.}
\label{fig:magfield-dependence2}
\end{figure}

\begin{figure}[!p]
\centering
\includegraphics[width=\columnwidth]{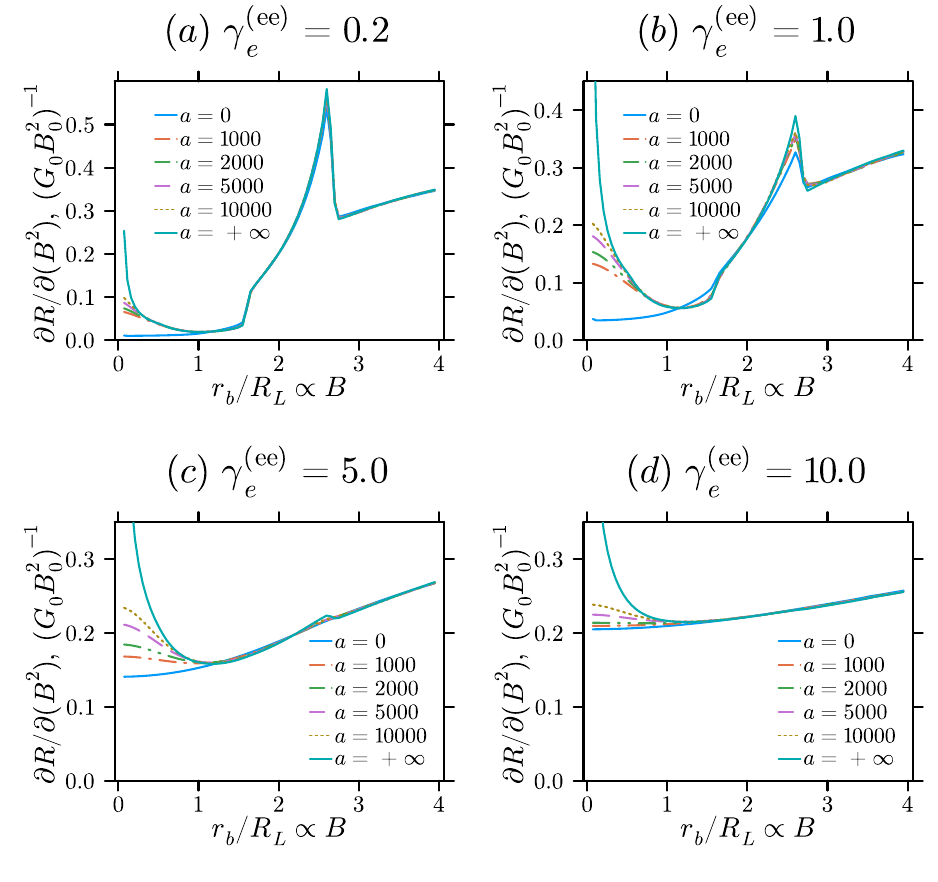}
\caption{Case $\gamma^{(\mathrm{mr})}=0.5$: Sensitivity $\partial R/\partial(B^2)$ of the resistance $R=1/G$ to the square of the magnetic field as a function of the inverse Larmor radius $r_b/R_L\propto B$ for different values of the phenomenological parameter $a$ and the even electron-electron scattering rate $\gamma_e^{(\mathrm{ee})}$.}
\label{fig:magfield-sensitivity2}
\end{figure}

\begin{figure}[!p]
\centering
\includegraphics[width=\columnwidth]{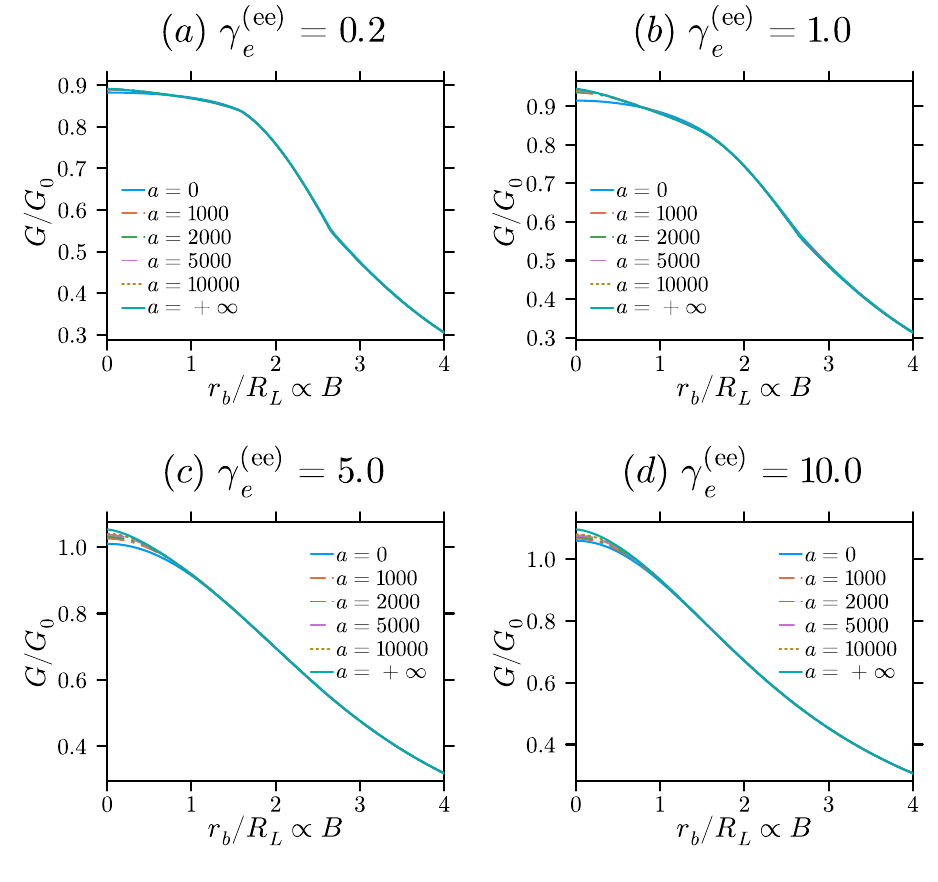}
\caption{Case $\gamma^{(\mathrm{mr})}=1.0$: Normalized total conductance $G/G_0$ of the Corbino disk as a function of the inverse Larmor radius $r_b/R_L\propto B$ for different values of the phenomenological parameter $a$ and the even electron-electron scattering rate $\gamma_{e}^{(\mathrm{ee})}$.}
\label{fig:magfield-dependence3}
\end{figure}

\begin{figure}[!p]
\centering
\includegraphics[width=\columnwidth]{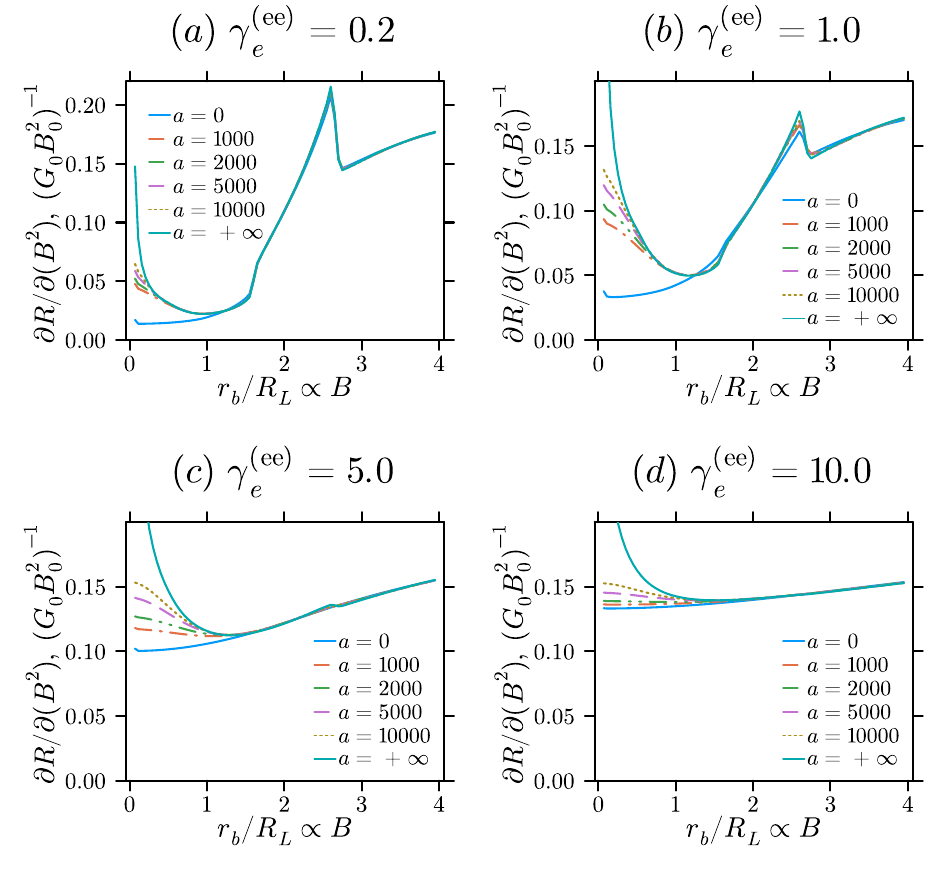}
\caption{Case $\gamma^{(\mathrm{mr})}=1.0$: Sensitivity $\partial R/\partial(B^2)$ of the resistance $R=1/G$ to the square of the magnetic field as a function of the inverse Larmor radius $r_b/R_L\propto B$ for different values of the phenomenological parameter $a$ and the even electron-electron scattering rate $\gamma_e^{(\mathrm{ee})}$.}
\label{fig:magfield-sensitivity3}
\end{figure}

Here, we present the results for the normalized conductance $G/G_0$ and resistance sensitivity $\partial R/\partial(B^2)$ for alternative values of $\gamma^{(\mathrm{mr})}$. Figures~\ref{fig:magfield-dependence2} and~\ref{fig:magfield-sensitivity2} display the normalized conductance and resistance sensitivity respectively for $\gamma^{(\mathrm{mr})}=0.5$, while Figs.~\ref{fig:magfield-dependence3} and~\ref{fig:magfield-sensitivity3} do the same for $\gamma^{(\mathrm{mr})}=1.0$.
\FloatBarrier

\clearpage


\end{document}